\documentclass[aps, prl, showpacs, amsfonts, amsmath, amssymb, twocolumn, floatfix, superscriptaddress, 10pt]{revtex4-2}
\usepackage[final]{graphicx}
\usepackage{xcolor}
\usepackage[colorlinks=true, linkcolor=blue, citecolor=blue, urlcolor=blue]{hyperref}
\usepackage{verbatim}
\usepackage{csquotes}
\usepackage{braket}
\usepackage{mathtools}
\usepackage[normalem]{ulem}

\newcommand{\vct}[1]{\mathbf{#1}}

\renewcommand\Re{\operatorname{Re}}
\renewcommand\Im{\operatorname{Im}}
\newcommand\Tr{\operatorname{Tr}}
\allowdisplaybreaks

\begin{document}


\title{Radiative heat transfer with a cylindrical waveguide decays logarithmically slow}

\author{Kiryl Asheichyk}
\email[]{asheichyk@bsu.by}
\affiliation{Department of Theoretical Physics and Astrophysics, Belarusian State University, 5 Babruiskaya Street, 220006 Minsk, Belarus}   
\author{Matthias Kr{\"u}ger}
\email[]{matthias.kruger@uni-goettingen.de}
\affiliation{Institute for Theoretical Physics, Georg-August-Universit{\"a}t G{\"o}ttingen, 37073 G{\"o}ttingen, Germany}

\date{\today}

\begin{abstract} 
Radiative heat transfer between two far-field-separated nanoparticles placed close to a perfectly conducting nanowire decays logarithmically slow with the interparticle distance. This makes a cylinder an excellent waveguide which  can transfer thermal electromagnetic energy to arbitrary large distances with almost no loss. It leads to a dramatic increase of the heat transfer, so that, for almost any (large) separation, the transferred energy can be as large as for isolated particles separated by a few hundred nanometers. A phenomenologically found analytical formula accurately describes the numerical results over a wide range of parameters.
\end{abstract}


\maketitle



Heat radiation (HR) and radiative heat transfer (HT) are very sensitive to changes in geometrical configuration and material properties when the system length scales are smaller or comparable to the thermal wavelength (roughly $ 8 \ \mu\textrm{m} $ at room temperature). This was first observed more than $ 50 $ years ago for the HT between two parallel plates, where the near-field HT shows a strong increase with decreasing the gap width due to the evanescent waves contribution, absent for far-field separations~\cite{Hargreaves1969, Polder1971}. Since then, researchers investigated near-field HR and HT in a variety of systems with objects of different shapes and materials, revealing plenty of interesting effects~\cite{Bimonte2017, Cuevas2018, Song2021, Biehs2021}.

An important question is whether these near-field effects can be propagated to the far field, thereby improving the efficiency of HT between objects at large separations. Recent studies found such a propagation possible in two cases: (i) for anisotropic objects with some of their dimensions smaller than the thermal wavelength~\cite{Fernandez-Hurtado2018_1, Fernandez-Hurtado2018_2, Thompson2018}; (ii) for objects that are placed in the proximity of intermediate objects~\cite{Saaskilahti2014, Messina2016, Asheichyk2017, Dong2018, Messina2018, Asheichyk2018, Zhang2019_1, Zhang2019_2, He2019}. In the first case, the far-field HT can greatly exceed the blackbody result, and for distances larger than the objects themselves, it decreases with the expected power law behavior~\cite{Fernandez-Hurtado2018_1, Fernandez-Hurtado2018_2, Thompson2018}.

The second case also provides a variety of interesting phenomena. The HT between two nanoparticles placed above a plate~\cite{Saaskilahti2014, Dong2018, Messina2018, Asheichyk2018, Zhang2019_1, He2019}, inside a two-plates cavity~\cite{Saaskilahti2014, Asheichyk2018}, or connected through the near field by a sphere~\citep{Asheichyk2017} can be enhanced by several orders of magnitude (compared to isolated particles) even for interparticle distances larger than the thermal wavelength. This enhancement and its mechanism strongly depend on the system geometry and material properties. A larger increase of the HT is achieved when the system supports resonant surface modes, for example, two SiC particles above a SiC plate~\cite{Saaskilahti2014, Dong2018, Messina2018, Asheichyk2018, Zhang2019_1, He2019} or inside a SiC cavity~\cite{Asheichyk2018}. However, as the interparticle distance grows, the effect is quickly diminished due to a strong absorption of the SiC plates~\cite{Saaskilahti2014, Dong2018, Messina2018, Asheichyk2018}. Less absorbing metallic plates provide a  longer ranged, but smaller effect~\cite{Dong2018, Asheichyk2018}. Using sophisticated structures can improve the efficiency, however only in a short range of far-field separations~\cite{Zhang2019_1, Zhang2019_2, He2019}. Is there a geometry that allows for a long range energy transport beyond the mentioned scales?

In this paper, we show that the HT between two far-field-separated nanoparticles placed in the proximity of a perfectly conducting cylinder is much larger than in other configurations studied before. It decays logarithmically as a function of the interparticle distance $ d $, compared to $ d^{-2} $ decay for the particles in vacuum~\cite{Volokitin2001} or $ d^{-1} $ for the particles placed in a metallic  cavity~\citep{Asheichyk2018}. As a consequence of the logarithmic decay, the ratio to the HT for isolated particles grows as $d^2$, e.g., exceeding it by 12 orders of magnitude with $d$ in the range of centimeters. Saying it differently, the HT at almost any large distance is comparable to the transfer between isolated particles at $d\approx500$~nm for the parameters studied. We analyze the dependence of the HT on the system parameters, providing a phenomenological analytical formula,  and discuss potential applications and implications of the observed phenomena.          

\begin{figure}[!t]
\begin{center}
\includegraphics[width=1.0\linewidth]{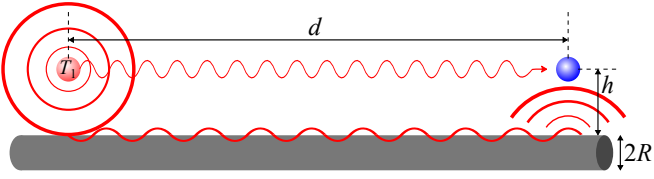}
\end{center}
\caption{\label{fig:System}Radiative heat transfer from particle $ 1 $ at temperature $ T_1 $ to particle $ 2 $ in the presence of an infinitely long perfectly conducting cylinder of radius $ R $. The near-field energy radiated by the first particle is captured by the cylinder and guided via surface waves to the second particle. These surface waves decay logarithmically slow with $ d $, resulting in a highly efficient heat transfer even for far-separated particles.}
\end{figure}

The considered system is depicted in Fig.~\ref{fig:System}. Two spherical particles are placed symmetrically in the proximity of an infinitely long perfectly conducting cylinder. We aim to compute the HT from particle $ 1 $ at temperature $ T_1 $ to particle $ 2 $. In general, there are other HT contributions, due to sources in the environment, the cylinder (if not a perfect conductor), and particle $2$~\cite{Kruger2012}. The studied contribution depends only on $T_1$, and  may be imagined, e.g., as the case with all temperatures except for $T_1$ equal to zero.

\begin{figure}[!t]
\begin{center}
\includegraphics[width=1.0\linewidth]{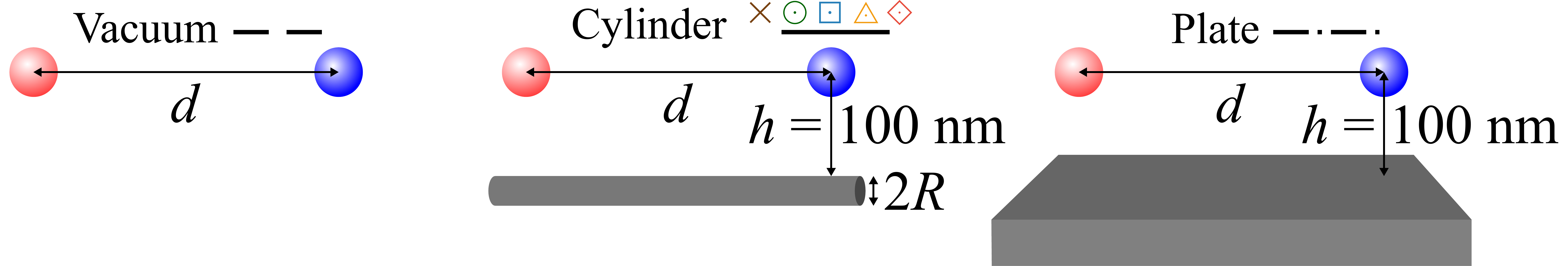}
\includegraphics[width=1.0\linewidth]{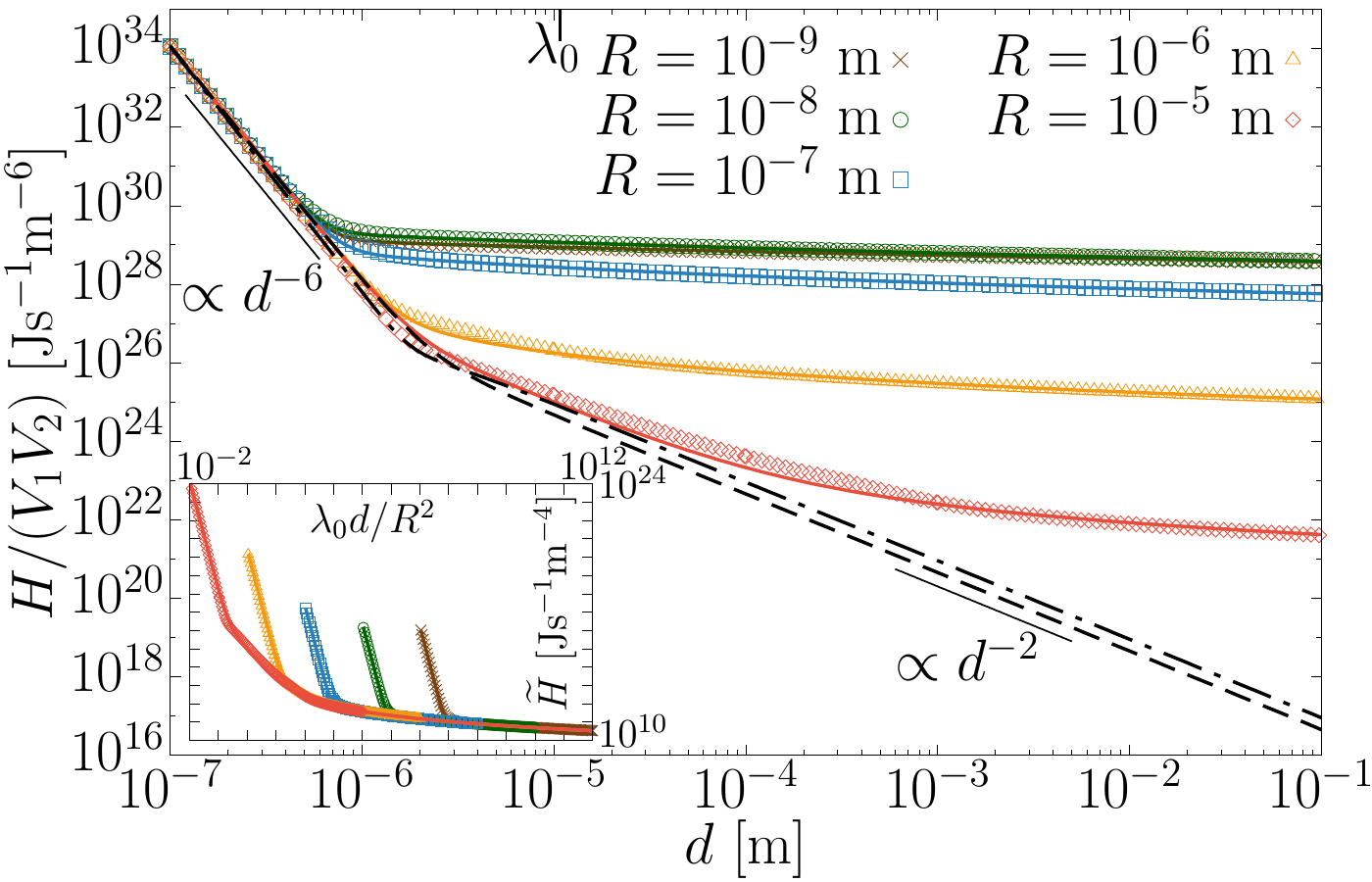}
\end{center}
\caption{\label{fig:HT_ddep}Heat transfer (normalized by particles' volumes) from SiC particle 1 at temperature $ T_1 = 300 \ \textrm{K} $ to SiC particle 2 in the presence of a perfectly conducting cylinder as a function of interparticle distance $ d $. The particles are placed symmetrically above the cylinder at distance $ h = 10^{-7} \ \textrm{m} $ from its surface (see the sketch). The results are given for different radii $ R $ of the cylinder and compared to cases of the particles in vacuum and above a perfectly conducting plate at the same $ h $. Points show numerically exact results (computed using Eq.~\eqref{eq:HT}, with numerically exact $ \Tr\left(\mathbb{G}\mathbb{G}^{\dagger}\right) $), while solid lines represent the approximate HT given by formula~\eqref{eq:HTanapprox}. The dominant wavelength $ \lambda_0 \approx 1.08 \times 10^{-5} \ \textrm{m} $. Inset shows rescaled curves, $ \widetilde{H} = H(R+h)^4/(V_1V_2\lambda_0^2) $ (same color coding), as a function of the rescaled distance $\lambda_0d/R^2$ (see main text below Eq.~\eqref{eq:HTanapprox}).}
\end{figure}

\begin{figure}[!t]
\begin{center}
\includegraphics[width=1.0\linewidth]{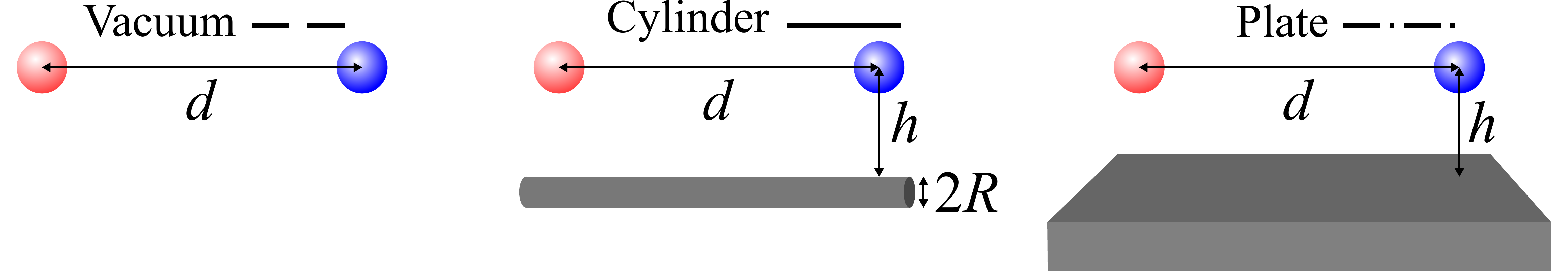}
\includegraphics[width=1.0\linewidth]{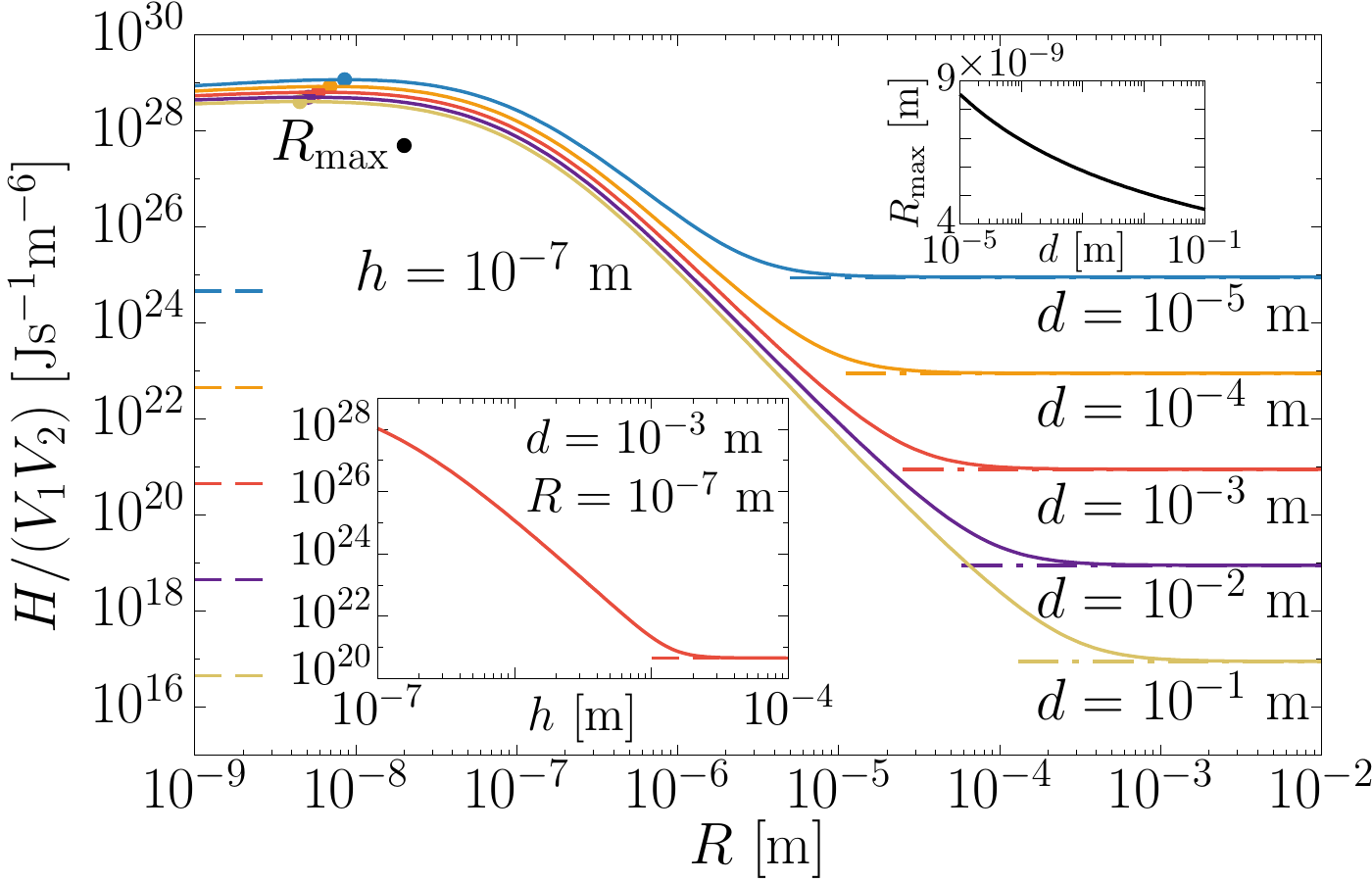}
\end{center}
\caption{\label{fig:HT_Rdep}Heat transfer (normalized by particles' volumes) from SiC particle 1 at temperature $ T_1 = 300 \ \textrm{K} $ to SiC particle 2 in the presence of a perfectly conducting cylinder as a function of its radius $ R $. The particles are placed symmetrically above the cylinder at distance $ h = 10^{-7} \ \textrm{m} $ from its surface, and the results are given for different interparticle distances $ d $ (see the sketch), using formula~\eqref{eq:HTanapprox}. Dashed and dashed-dotted lines give the corresponding heat transfer in vacuum and in the presence of a perfectly conducting plate (with the same $ h $), respectively. The left inset shows $H$ as a function of $ h $ for $ d = 10^{-3} \ \textrm{m} $ and $ R = 10^{-7} \ \textrm{m} $, with the vacuum case, approached for large $h$, shown as a dashed line. The right inset shows $ R_{\textrm{max}} $ for $ h = 10^{-7} \ \textrm{m} $ (the radius which maximizes the HT) as a function of $ d $.}
\end{figure}

\begin{figure}[!t]
\begin{center}
\includegraphics[width=1.0\linewidth]{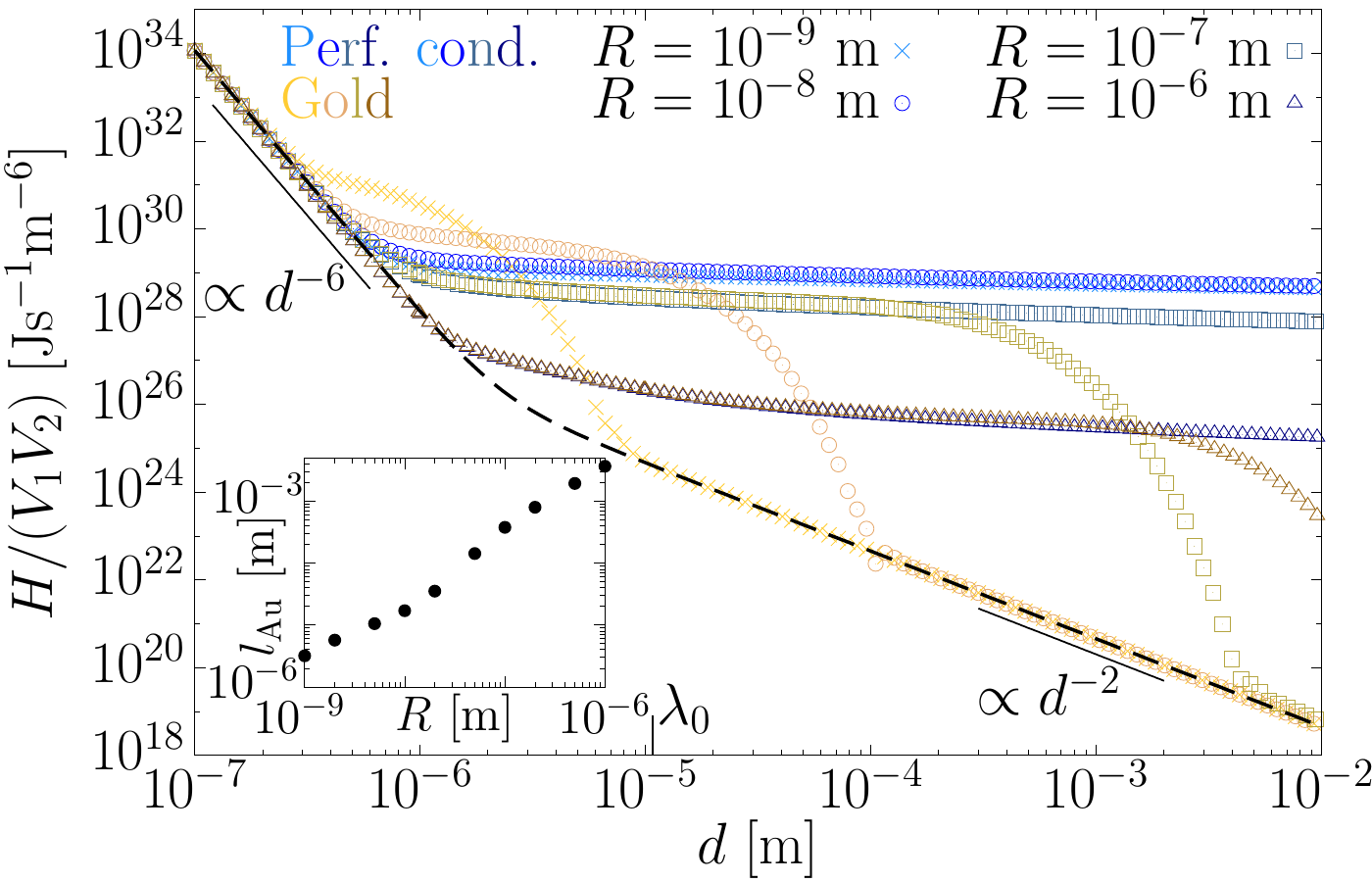}
\end{center}
\caption{\label{fig:HT_gold}Heat transfer from SiC particle 1 to SiC particle~2 as a function of $ d $, with all parameters as in Fig.~\ref{fig:HT_ddep}. Additionally to the perfectly conducting cylinder (blue shaded points) we show a gold cylinder (shades of gold). Inset shows the decay length $l_{\textrm{Au}}$ (see Supplemental Material~\cite{Note2} for definition) for a gold cylinder, as a function of $ R $, characterizing the transition of the HT from slow to fast decay.}
\end{figure}

Aiming at a proof of concept, and to simplify the problem, we use the point particle (PP) limit, where the two particles are small compared to the thermal wavelength, the particle's skin depth, and distances $ d $ and $ h $~\cite{Asheichyk2017, Asheichyk2018}~\footnote{The particles are assumed to be nonmagnetic, i.e., their magnetic permeabilities equal unity.}. The results in Figs.~\ref{fig:HT_ddep},~\ref{fig:HT_Rdep}, and~\ref{fig:HT_gold} are valid for particle radii $R_i\ll h$, i.e.,  $R_i \approx 10 \ \textrm{nm} $ or smaller for the given value of $ h = 100 \ \textrm{nm} $.

Problems of HR and HT in many-body systems can be studied within frameworks of fluctuational electrodynamics~\cite{Rytov1958, Rytov1989} and scattering theory~\cite{Ben-Abdallah2011, Messina2014, Muller2017}. For our system, the HT from PP $ 1 $ to PP $ 2 $ reads as~\cite{Asheichyk2017, Ben-Abdallah2011}
\begin{equation}
H = \frac{32\pi\hbar}{c^4} \int_0^\infty d\omega \frac{\omega^5}{e^{\frac{\hbar\omega}{k_{\textrm{B}}T_1}}-1}\Im(\alpha_1)\Im(\alpha_2)\Tr\left(\mathbb{G}\mathbb{G}^{\dagger}\right),
\label{eq:HT}
\end{equation}
where $ c $ is the speed of light in vacuum, and $ \hbar $ and $ k_{\textrm{B}} $ are Planck's and Boltzmann's constants, respectively. $ \Tr\left(\mathbb{G}\mathbb{G}^{\dagger}\right) $ is the trace of the matrix product of the dyadic Green's function (GF) $\mathbb{G}$ of the cylinder, evaluated at the particles' coordinates, and its conjugate transpose $ \mathbb{G}^{\dagger}$. The GF encodes the system geometry (it is a function of $ R $, $ h $, and $ d $), and hence determines the role of the cylinder in the HT. Note that it is also a function of the wave number $ k = \omega/c $ (i.e., the absolute value of the wave vector) or the corresponding wavelength $ \lambda = 2\pi/k $.
\begin{equation}
\alpha_i (\omega)= \frac{\varepsilon_i(\omega)-1}{\varepsilon_i(\omega)+2}R_i^3
\label{eq:polarizability}
\end{equation}
is the electrical dipole polarizability of particle $ i $, with $ R_i $ and $ \varepsilon_i $ being the radius and the frequency-dependent dielectric permittivity, respectively. Polarizabilities determine the radiation and absorption strength of the particles. For numerical illustration, we use both particles to be made of SiC, $ \varepsilon_1 = \varepsilon_2 = \varepsilon_{\textrm{SiC}} $, where~\cite{Spitzer1959}
\begin{equation}
\varepsilon_{\textrm{SiC}}(\omega) = \varepsilon_\infty\frac{\omega^2-\omega_{\textrm{LO}}^2+i\omega\gamma}{\omega^2-\omega_{\textrm{TO}}^2+i\omega\gamma},
\label{eq:epsilon_SiC}
\end{equation}
with $ \varepsilon_\infty=6.7 $, $ \omega_{\textrm{LO}}=1.82\times10^{14} \ \textrm{rad} \ \textrm{s}^{-1} $, $ \omega_{\textrm{TO}}=1.48\times10^{14} \ \textrm{rad} \ \textrm{s}^{-1} $, $ \gamma=8.93\times10^{11} \ \textrm{rad} \ \textrm{s}^{-1} $.

For details of the GF, we refer the reader to Supplemental Material~\footnote{See Supplemental Material for the Green's function of a cylinder, dependence of $ \Tr\left(\mathbb{G}\mathbb{G}^{\dagger}\right) $ on the system parameters, details of approximation~\eqref{eq:Tr}, comparison between Eqs.~\eqref{eq:HTanapproxGF} and~\eqref{eq:HTanapprox}, details of the ratio $ H/H_{\textrm{total}} $, and the heat transfer in the presence of a gold cylinder.}. It is worth noticing that the GF of a cylinder contains both integration over a certain component of the wave vector and summation over multipoles~\cite{Golyk2012, Note2}, in contrast to the GF of a plate (only integration)~\cite{Asheichyk2017, Dong2018, Messina2018, Zhang2019_1, Zhang2019_2, He2019, Nikitin2013} or a sphere (only summation)~\cite{Asheichyk2017}. This makes the numerical computations difficult requiring long computation times.

As for the system parameters, we fix $ T_1 = 300 \ \textrm{K} $, such that the corresponding thermal and dominant wavelengths are $ \lambda_{T_1} \approx 7.63 \times 10^{-6} \ \textrm{m} $ and $ \lambda_0 \approx 1.08 \times 10^{-5} \ \textrm{m} $, respectively, the latter resulting from the material properties of Eq.~\eqref{eq:epsilon_SiC} and the polarizability in Eq.~\eqref{eq:polarizability}. The particles are placed at a small near-field distance above the cylinder, $ h = 10^{-7} \ \textrm{m} $, for a strong coupling between particles and cylinder. The dependence on $ h $ is discussed below and in Supplemental Material~\cite{Note2}. With $ d \agt 10^{-7} \ \textrm{m} $, the PP limit of Eq.~\eqref{eq:HT} is valid for   $ R_i \lesssim 10^{-8} \ \textrm{m} $.
Because $ H \propto V_1V_2 $, with $ V_i $ being the volume of particle $ i $ (see Eq.~\eqref{eq:HT}), we do not give $ R_i $ explicitly and normalize the HT by $ V_1V_2 $.

Figure \ref{fig:HT_ddep} shows the HT as a function of interparticle distance $ d $. We note a drastic effect of the presence of the cylinder on the far-field HT. For distances larger than a few hundred nanometers, the HT seems to be almost {\it independent} of distance $d$. This yields a strong enhancement over the vacuum HT or the case of a close-by plate, the more so, the larger $d$:  For example, for $R = 10^{-8} \ \textrm{m} $ and $ d = 10^{-1} \ \textrm{m} $, the HT is larger than the vacuum HT by $ 12 $ orders of magnitude.  In other words, the HT between the particles placed above of a wire and separated by $ 10 $ centimeters is the same as the HT between isolated particles which are just a few hundred nanometers apart. We attribute this to the system geometry, i.e., the cylinder guides the energy in the preferred direction via surface waves.

How does this effect depend on $R$? In Fig.~\ref{fig:HT_ddep}, we note that it is especially strong for a thin cylinder, and the enhancement decreases the thicker the cylinder. For large $R$, the HT approaches the result of two particles close to a plate as may be expected~\footnote{For $ d \gg \{\lambda_0, \ h\} $, the plate HT scales the same as the vacuum HT, i.e., $ \sim d^{-2} $, but the former is two times larger~\cite{Note2}.}. The strong effect of a thin cylinder may be related to its ability to concentrate the near-field energy radiated by particle $ 1 $ over a smaller surface area, such that the energy loss of the surface waves traveling along the cylinder to particle $ 2 $ is minimized. Figure~\ref{fig:HT_Rdep} shows the result as a function of $R$, for various fixed values of $d$, displaying the mentioned approach of the plate result for large $R$, and the general decrease with $R$. However, for very small $R$, the HT again decreases, so that a maximal value for $R$ appears.

Can we characterize the observed behavior analytically? Looking at Fig.~\ref{fig:HT_ddep}, one may expect that the HT decays slower than any power law. Therefore, we presumed a logarithmic dependence of $ \Tr\left(\mathbb{G}\mathbb{G}^{\dagger}\right) $ for an analytical approximation of the numerical data~\footnote{Logarithmic dependence on the system parameters is a known feature of a cylinder. It was observed in the heat radiation of a cylinder~\cite{Golyk2012} as well as in Casimir forces involving cylinders~\cite{Rahi2009, Noruzifar2011}.}, which leads to the scaling of the HT in Eq.~\eqref{eq:HTanapprox}. Indeed, considering a logarithmic decay with $ d $, as well as the dependence on $ R $ and $ h $, we found that $ \Tr\left(\mathbb{G}\mathbb{G}^{\dagger}\right) $ can be well approximated by ($k=\omega/c$)
\begin{align}
\notag \Tr\left(\mathbb{G}\mathbb{G}^{\dagger}\right) \approx \ & \frac{1}{8\pi^2 d^2}\left[1+\frac{1}{k^2d^2}+\frac{3}{k^4d^4}\right]\\
& + \frac{1}{4\pi^2k^2(R+h)^4\ln^2\left[1+\frac{\sqrt{2}\sqrt{d^2+4h^2}}{kR^2}\right]},
\label{eq:Tr}
\end{align}
where the first term is the vacuum part~\cite{Asheichyk2017}, while the second term is the cylinder contribution. Formula~\eqref{eq:Tr} is discussed in detail in Supplemental Material~\cite{Note2}, where we show that it is  a good approximation for almost any regimes of parameters. In case $ h \ll \lambda \lesssim d $ (i.e., what we are interested in), the agreement with the numerical result is very good, and it becomes excellent when we also have $ \lambda d \gg 16R^2 $ or $ \lambda d \ll 4R^2 $. These two conditions correspond to the cylinderlike or platelike limits of $ \Tr\left(\mathbb{G}\mathbb{G}^{\dagger}\right) $, respectively~\cite{Note2}. Substituting Eq.~\eqref{eq:Tr} into Eq.~\eqref{eq:HT}, one gets for the heat transfer
\begin{align}
\notag H \approx \ & \frac{4\hbar}{\pi c^4} \int_0^\infty d\omega \frac{\omega^5}{e^{\frac{\hbar\omega}{k_{\textrm{B}}T_1}}-1}\Im(\alpha_1)\Im(\alpha_2)\\
\notag & \times \Bigg\{\frac{1}{d^2}\left[1+\frac{c^2}{\omega^2d^2}+\frac{3c^4}{\omega^4d^4}\right]\\
& \ \ \ \ + \frac{2c^2}{\omega^2(R+h)^4\ln^2\left[1+\frac{\sqrt{2}c\sqrt{d^2+4h^2}}{\omega R^2}\right]}\Bigg\}.
\label{eq:HTanapproxGF}
\end{align}

Formula~\eqref{eq:HTanapproxGF} can be further simplified. For SiC PPs and $ T_1 = 300 \ \textrm{K} $, the HT spectrum is strongly peaked at $ \omega_0 = 1.75194 \times 10^{14} \ \textrm{rad} \ \textrm{s}^{-1} $. It is the resonance frequency of SiC PP (equivalently, the resonance frequency of $ \alpha_i $ in Eq.~\eqref{eq:polarizability}). The corresponding dominant wavelength $ \lambda_0 \approx 2\pi c/\omega_0 \approx 1.08 \times 10^{-5} \ \textrm{m} $ is close to $ \lambda_{T_1} $. Since the cylinder is a perfect conductor, $ \Tr\left(\mathbb{G}\mathbb{G}^{\dagger}\right) $ has no resonances. Therefore, we can replace $ \omega $ with $ \omega_0 $ in the trace and pull the trace out off the frequency integral. As a result, we finally obtain
\begin{align}
\notag H \approx \ & \frac{4\hbar}{\pi c^4}\Bigg\{\frac{1}{d^2}\left[1+\frac{c^2}{\omega_0^2d^2}+\frac{3c^4}{\omega_0^4d^4}\right]\\
\notag & + \frac{2c^2}{\omega_0^2(R+h)^4\ln^2\left[1+\frac{\sqrt{2}c\sqrt{d^2+4h^2}}{\omega_0R^2}\right]}\Bigg\}\\
& \ \ \ \ \times \int_0^\infty d\omega \frac{\omega^5}{e^{\frac{\hbar\omega}{k_{\textrm{B}}T_1}}-1}\Im(\alpha_1)\Im(\alpha_2),
\label{eq:HTanapprox}
\end{align}
which agrees almost perfectly with Eq.~\eqref{eq:HTanapproxGF}, as shown in Supplemental Material~\cite{Note2}, and provides a very good approximation for the exact HT, as can be seen in Fig.~\ref{fig:HT_ddep}: Its inset shows that $\widetilde{H} = H(R+h)^4/(V_1V_2\lambda_0^2)$, plotted as a function of $\lambda_0d/R^2$, leads to a collapse for large $d$ of the curves for all parameters shown. Formula~\eqref{eq:HTanapprox} is the main result of this paper. It states that the HT in the presence of a cylinder decays logarithmically with $ d $ for large $ d $, small $ h $, and small $ R $. More specifically, if $d \gtrsim \lambda_0$, the logarithmic behavior dominates when $ h \ll \lambda_0  $ and $ \lambda_0 d \gg 16R^2 $~\cite{Note2}. Note that, for a given $ d $ and for $ R \ll h $, the HT is a very strong function of $ h $, i.e., it scales as $ h^{-4} $. For $ h \ll R $, the HT approaches an $h$ independent value.

The logarithmic decay in Eq.~\eqref{eq:HTanapprox} suggests that a cylinder is an excellent waveguide for the HT. We are not aware that any other {\it unconfined} geometry can outperform the one in Fig.~\ref{fig:System} in terms of the HT efficiency for large interparticle distances. Using formula~\eqref{eq:HTanapprox}, the curves in Fig.~\ref{fig:HT_ddep} can be prolonged to arbitrarily large $ d $. Imagine a thought experiment with a nanoparticle being in Minsk, while the other is in G\"ottingen, i.e., $ d \approx 1200 \ \textrm{km} $, both placed close to an ideal nanowire ($ h = R = 100 \ \textrm{nm} $). According to Eq.~\eqref{eq:HTanapprox}, the HT measured in such a thought experiment is the same as the HT between isolated particles separated by $ d \approx 1.5 \ \mu\textrm{m} $. Roughly speaking,~\enquote{the logarithm does not care} whether the distance is of the order of a millimeter, kilometer, or the size of the Earth.

Noting that the curves in Fig.~\ref{fig:HT_Rdep} are computed using Eq.~\eqref{eq:HTanapprox}, they can be discussed in more detail. Interestingly, there is an optimal value of the radius, $ R_{\textrm{max}} $, where the HT has a maximum. This maximum slightly shifts to a smaller $ R $ with increase of $ d $, and it lays around $ R = 5 \times 10^{-9} \ \textrm{m} $ (see the right inset). For all considered $ d $, there is a slow logarithmic growth until $ R \ll h $ (encoded in the $ R $ dependence of the logarithm) followed by a fast decay ($ \sim (R+h)^{-4} $) which finally bends over to the plate result. This convergence to the plate HT occurs when $ 4R^2 \gg \lambda_0d $, i.e., it shifts to a larger $ R $ with increase of $ d $~\cite{Note2}. Expression for the HT in the plate limit can be straightforwardly obtained from Eq.~\eqref{eq:HTanapprox} using a large $ R $ expansion of the logarithm~\cite{Note2}. As Fig.~\ref{fig:HT_ddep}, Fig.~\ref{fig:HT_Rdep} reveals that a larger enhancement of the HT occurs for larger $ d $ and a thin cylinder. The left inset gives the HT as a function of $ h $, which shows a strong increase with placing the particles closer to a cylinder. Note that this increase saturates once $ h $ becomes comparable to $ R $. In Supplemental Material~\cite{Note2}, we give the detailed $ R $ and $ h $ dependence of both numerically exact and approximated $ \Tr\left(\mathbb{G}\mathbb{G}^{\dagger}\right) $ for both near- and far-field $ d $.

As seen in Fig.~\ref{fig:HT_ddep} and Eq.~\eqref{eq:HTanapproxGF}, the HT in the presence of a cylinder follows the vacuum result until a certain distance, where it bends away and is nearly $d$ independent. The HT with the cylinder for nearly any large $ d $ is thus comparable to the HT between isolated particles separated by a much smaller distance $d_{\textrm{zoom}}$. In other words, the cylinder allows to \enquote{zoom in} from almost any large distance to  $d_{\textrm{zoom}}$. In the case of 
$d_{\textrm{zoom}} \ll \lambda_0$, it reads as (found from formula~\eqref{eq:HTanapprox}~\cite{Note2})
\begin{equation}
d_{\textrm{zoom}} \approx \left[\lambda_0^2(R+h)^4\right]^{\frac{1}{6}}.
\label{eq:HT_dzoom}
\end{equation}
This zooming in is performed by bringing the particles close to the cylinder down to distance $h$.

What about imperfect conductors? For a real conductor, the surface waves traveling along the cylinder are damped, and the logarithmic decay is eventually cut off (see Fig.~\ref{fig:HT_gold} and Supplemental Material~\cite{Note2} for details). For a gold cylinder this occurs at roughly a length of $4000 R$, i.e., in the millimeter range for $R \approx 100 \ \textrm{nm}$, so that a gold microwire enhances the HT for over millimeter distances by more than 5 orders of magnitude. The amplitude of the HT generally increases for real materials, a study which we leave for future work.

Finally, it is worth mentioning that, despite the large far-field HT in the presence of a cylinder, the transferred energy is for large $ d $ small compared to the total energy emitted by particle 1. Within the PP limit, the maximum ratio between transferred and emitted energy is around $10^{-4} $, see Supplemental Material~\cite{Note2}. This is partly due to the observation that also the emitted energy itself increases strongly in the presence of a cylinder, as will be investigated in future work. 

A cylindrical waveguide is an excellent tool to efficiently transfer thermal energy between far-separated objects. The HT efficiency with a cylinder present can be more than $ 10 $ orders of magnitude better than for isolated objects, i.e., in this aspect much better than the efficiency with other waveguide geometries~\cite{Saaskilahti2014, Messina2016, Asheichyk2017, Dong2018, Messina2018, Asheichyk2018, Zhang2019_1, Zhang2019_2, He2019}. This effect can be applied in a variety of setups, e.g., for efficient far-field cooling or heating. These findings may also drastically affect many-body effects, promising strong nonadditivity of the HT~\cite{Cuevas2018, Song2021, Biehs2021}, for example, using more than one cylinder or more than two particles. A cylinder can also greatly influence the thermalization in complex setups~\cite{Kruger2012, Tschikin2012, Yannopapas2013, Messina2013, Dyakov2014, Nikbakht2015, Reina2021} and may drastically alter the heat transfer eigenmodes~\cite{Sanders2021}, where the (dominant) cylinder eigenmode is expected.
A highly directional energy transport with a cylinder can greatly affect heat transfer diffusion~\cite{Ben-Abdallah2013, Latella2018}. Using cylinders with nonreciprocal material may strongly improve rectification properties~\cite{Cuevas2018, Song2021}. 

As for an experimental realization of the studied system, we propose using two atomic force microscope (AFM) tips \cite{Narayanaswamy2008, Shen2009, Rousseau2009} placed close to a wire.

\vspace{7pt}
\begin{acknowledgments}
This research was funded by the Deutsche Forschungsgemeinschaft (DFG, German Research Foundation) through the Walter Benjamin fellowship (Project No. 453458207). 
\end{acknowledgments}



%



\onecolumngrid
\clearpage
\section{Supplemental material}
\setcounter{equation}{0}
\renewcommand{\theequation}{S\arabic{equation}}
\renewcommand{\theHequation}{S\arabic{equation}}
\setcounter{figure}{0}
\renewcommand{\thefigure}{S\arabic{figure}}
\renewcommand{\theHfigure}{S\arabic{figure}}

\subsection{Green's function of an infinitely long perfectly conducting cylinder}
We consider an infinitely long cylinder of radius $ R $ (see Fig.~\ref{fig:System} in the main text) and work in cylindrical coordinate system $ (r, \varphi, z) $, where the $ z $ axis coincides with the symmetry axis of the cylinder. We aim to find the GF $ \mathbb{G} = \mathbb{G}(\vct{r}, \vct{r}') $ of the cylinder, with both spatial arguments $ \vct{r} $ and $ \vct{r}' $ lying outside the cylinder. The GF can be separated into two parts~\cite{Bimonte2017, Asheichyk2017, Kruger2012, Muller2017, Golyk2012, Rahi2009}:
\begin{equation}
\mathbb{G} = \mathbb{G}_0 + \mathbb{G}_{\mathbb{T}},
\label{eq:G_separation}
\end{equation}
where $ \mathbb{G}_0 $ is the free space GF and $ \mathbb{G}_{\mathbb{T}} = \mathbb{G}_0\mathbb{T}\mathbb{G}_0 $ (here, the operator multiplication is understood~\cite{Bimonte2017, Asheichyk2017, Kruger2012, Muller2017, Rahi2009}) is the scattering part associated with the scattering operator $ \mathbb{T} $ of the cylinder~\cite{Golyk2012}. For the information about electromagnetic operators, we refer the reader to Refs.~\cite{Bimonte2017, Asheichyk2017, Kruger2012, Muller2017, Rahi2009}. 

The free GF is well known in closed form, but it is typically given in Cartesian coordinate system~\cite{Song2021, Biehs2021, Saaskilahti2014, Asheichyk2017, Dong2018, Messina2018, Zhang2019_1, He2019}. In order to obtain the free GF in cylindrical coordinates, one has to apply the corresponding transformation~\cite{Nikitin2013}.

The scattering part can be written as an expansion in outgoing cylindrical waves~\cite{Golyk2012}:
\begin{equation}
\mathbb{G}_{\mathbb{T}} = \frac{i}{8\pi}\sum_{P,P'}\sum_{n=-\infty}^{\infty}(-1)^n\int_{-\infty}^{\infty}dk_z\vct{P}^{\textrm{out}}_{n,k_z}(\vct{r})\otimes\vct{P}^{'\textrm{out}}_{-n,-k_z}(\vct{r}')T_{n,k_z}^{PP'}.
\label{eq:GT_expansion}
\end{equation}
Here, $ P, \ P' = \{M, N\} $ denote polarization ($ M $ for magnetic and $ N $ for electric), $ n \in \mathbb{Z} $ represents the multipole order, and $ k_z $ is the $ z $ component of the wave vector; symbol $ \otimes $ denotes the dyadic product. The waves $ \vct{M}^{\textrm{out}}_{n,k_z}$ and $ \vct{N}^{\textrm{out}}_{n,k_z}$ are given in Ref.~\cite{Golyk2012}. The scattering matrix elements $ T_{n,k_z}^{PP'} $, corresponding to the scattering operator $ \mathbb{T} $, depend on the radius and material of the cylinder~\cite{Golyk2012, Rahi2009, Noruzifar2011}. In case of a perfectly conducting cylinder, they simplify to~\cite{Rahi2009}
\begin{subequations}
\begin{alignat}{1}
& T_{n,k_z}^{MM} = -\frac{J'_n(qR)}{H'_n(qR)},\label{eq:TMM}\\
& T_{n,k_z}^{NN} = -\frac{J_n(qR)}{H_n(qR)},\label{eq:TNN}\\
& T_{n,k_z}^{MN} = T_{n,k_z}^{NM} = 0,\label{eq:TMN}
\end{alignat}
\end{subequations}
where $ J_n $ and $ H_n $ are the Bessel function and the Hankel function of the first kind of order $ n $, respectively, $ q = \sqrt{k^2-k_z^2} $, and $ J'_n(qR) \equiv \frac{dJ_n(qR)}{d(qR)} $.

For the geometry in Fig.~\ref{fig:System} in the main text, $ r = r' = R + h $, $ \varphi = \varphi' $, and $ z' - z = d $. Such a symmetric configuration allows to greatly simplify the GF. $ \mathbb{G}_0 $ is diagonal (with $ G_{011} = G_{022} $) and depends only on $ d $. Using cylindrical waves from Ref.~\cite{Golyk2012} and substituting Eqs.~\eqref{eq:TMM},~\eqref{eq:TNN}, and~\eqref{eq:TMN} into Eq.~\eqref{eq:GT_expansion}, we find that
\begin{equation}
\mathbb{G}_{\mathbb{T}} = 
\begin{pmatrix}
G_{\mathbb{T}11} & 0 & G_{\mathbb{T}13}\\
0 & G_{\mathbb{T}22} & 0\\
-G_{\mathbb{T}13} & 0 & G_{\mathbb{T}33}
\end{pmatrix},
\label{eq:GT_matrix}
\end{equation}
with
\begin{subequations}
\begin{alignat}{4}
\notag & G_{\mathbb{T}11} = -\frac{i}{4\pi}\int_0^{\infty}dk_z\frac{k_z^2}{k^2}\frac{H^2_1(qr)J_0(qR)}{H_0(qR)}\cos(k_zd)\\
& \ \ \ \ \ \ \ \ \ -\frac{i}{2\pi}\sum_{n=1}^{\infty}\int_0^{\infty}dk_z\left[\frac{n^2}{(qr)^2}\frac{H^2_n(qr)J'_n(qR)}{H'_n(qR)} + \frac{k_z^2}{k^2}\frac{\left[H'_n(qr)\right]^2J_n(qR)}{H_n(qR)}\right]\cos(k_zd),\label{eq:GT11}\\
\notag & G_{\mathbb{T}22} = -\frac{i}{4\pi}\int_0^{\infty}dk_z\frac{H^2_1(qr)J_1(qR)}{H_1(qR)}\cos(k_zd)\\
& \ \ \ \ \ \ \ \ \ -\frac{i}{2\pi}\sum_{n=1}^{\infty}\int_0^{\infty}dk_z\left[\frac{[H'_n(qr)]^2J'_n(qR)}{H'_n(qR)} +\frac{n^2k_z^2}{k^2(qr)^2}\frac{H^2_n(qr)J_n(qR)}{H_n(qR)}\right]\cos(k_zd),\label{eq:GT22}\\
& G_{\mathbb{T}33} = -\frac{i}{4\pi}\int_0^{\infty}dk_z\frac{q^2}{k^2}\frac{H^2_0(qr)J_0(qR)}{H_0(qR)}\cos(k_zd) -\frac{i}{2\pi}\sum_{n=1}^{\infty}\int_0^{\infty}dk_z\frac{q^2}{k^2}\frac{H^2_n(qr)J_n(qR)}{H_n(qR)}\cos(k_zd),\label{eq:GT33}\\
& G_{\mathbb{T}13} = \frac{i}{4\pi}\int_0^{\infty}dk_z\frac{qk_z}{k^2}\frac{H_0(qr)H_1(qr)J_0(qR)}{H_0(qR)}\sin(k_zd) -\frac{i}{2\pi}\sum_{n=1}^{\infty}\int_0^{\infty}dk_z\frac{qk_z}{k^2}\frac{H_n(qr)H'_n(qr)J_n(qR)}{H_n(qR)}\sin(k_zd).\label{eq:GT13}
\end{alignat}
\end{subequations}
Note that $ \mathbb{G}_{\mathbb{T}} $ is a function of $ R $, $ h $, and $ d $. In numerical computations of $ \mathbb{G}_{\mathbb{T}} $, appropriate cutoffs, depending on these parameters and replacing infinity in the upper integration and summation limits, have to be chosen.

\subsection{Numerical results and analytical approximation for \texorpdfstring{$ \Tr\left(\mathbb{G}\mathbb{G}^{\dagger}\right) $}{the trace}}

\subsubsection{The dependence on the system parameters}
Since the geometry of the HT between PPs is fully determined by the GF, more precisely, by $ \Tr\left(\mathbb{G}\mathbb{G}^{\dagger}\right) $ (see Eq.~\eqref{eq:HT} in the main text), the dependence of these two quantities on geometrical parameters are expected to be similar, allowing to make qualitative predictions for the HT without performing the frequency integral in Eq.~\eqref{eq:HT} in the main text. The HT spectrum is strongly peaked at frequencies close to the dominant frequency of SiC PP radiation at $ T_1 = 300 \ \textrm{K} $, $ \omega_0 = 1.75194 \times 10^{14} \ \textrm{rad} \ \textrm{s}^{-1} $, with the corresponding wave number $ k_0 = \frac{\omega_0}{c} $ and wavelength $ \lambda_0 =  \frac{2\pi}{k_0} $. Therefore, $ \Tr\left(\mathbb{G}\mathbb{G}^{\dagger}\right) $ should be evaluated at $ k = k_0 $ to predict the behavior of the HT.

\begin{figure}[!b]
\begin{center}
\includegraphics[width=0.7\linewidth]{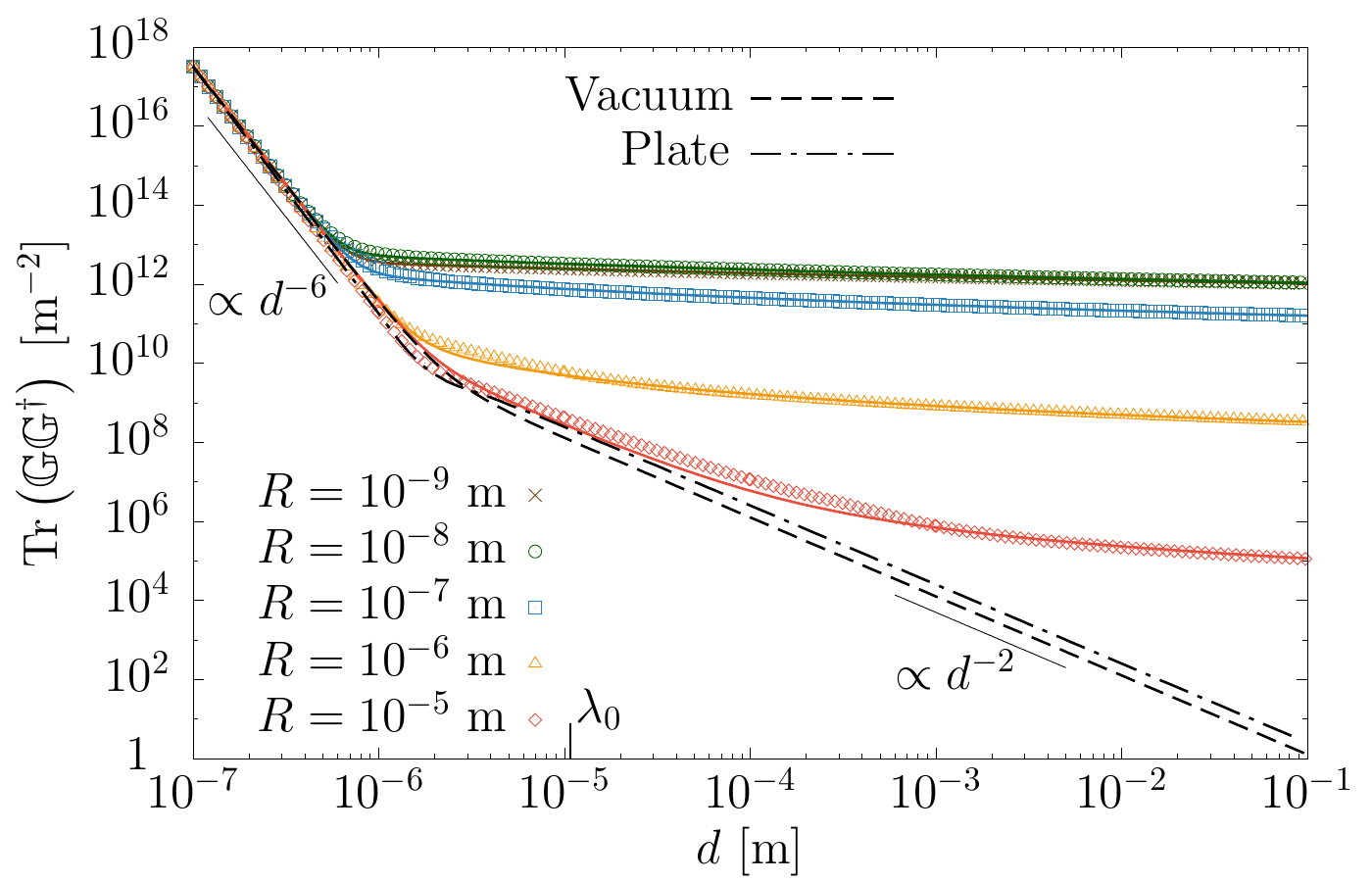}
\end{center}
\caption{\label{fig:GF_ddep}$ \Tr\left(\mathbb{G}\mathbb{G}^{\dagger}\right) $ of a perfectly conducting cylinder evaluated at wave number $ k_0 $ (the dominant wave number of SiC PP radiation) as a function of distance $ d $ between PPs. The particles are placed symmetrically above the cylinder at distance $ h = 10^{-7} \ \textrm{m} $ from its surface (see Fig.~\ref{fig:System} in the main text). The results are given for different radii $ R $ of the cylinder and compared to cases of PPs in vacuum and above a perfectly conducting plate at the same $ h $. Points show numerically exact results (i.e., $ \mathbb{G}_{\mathbb{T}} $ is computed using Eqs.~\eqref{eq:GT_matrix},~\eqref{eq:GT11},~\eqref{eq:GT22},~\eqref{eq:GT33}, and~\eqref{eq:GT13}), while solid lines represent analytical approximation given by Eq.~\eqref{eq:Tr} in the main text. $ \lambda_0 = \frac{2\pi}{k_0} $.}
\end{figure}

\begin{figure}[!t]
\begin{center}
\includegraphics[width=0.497\linewidth]{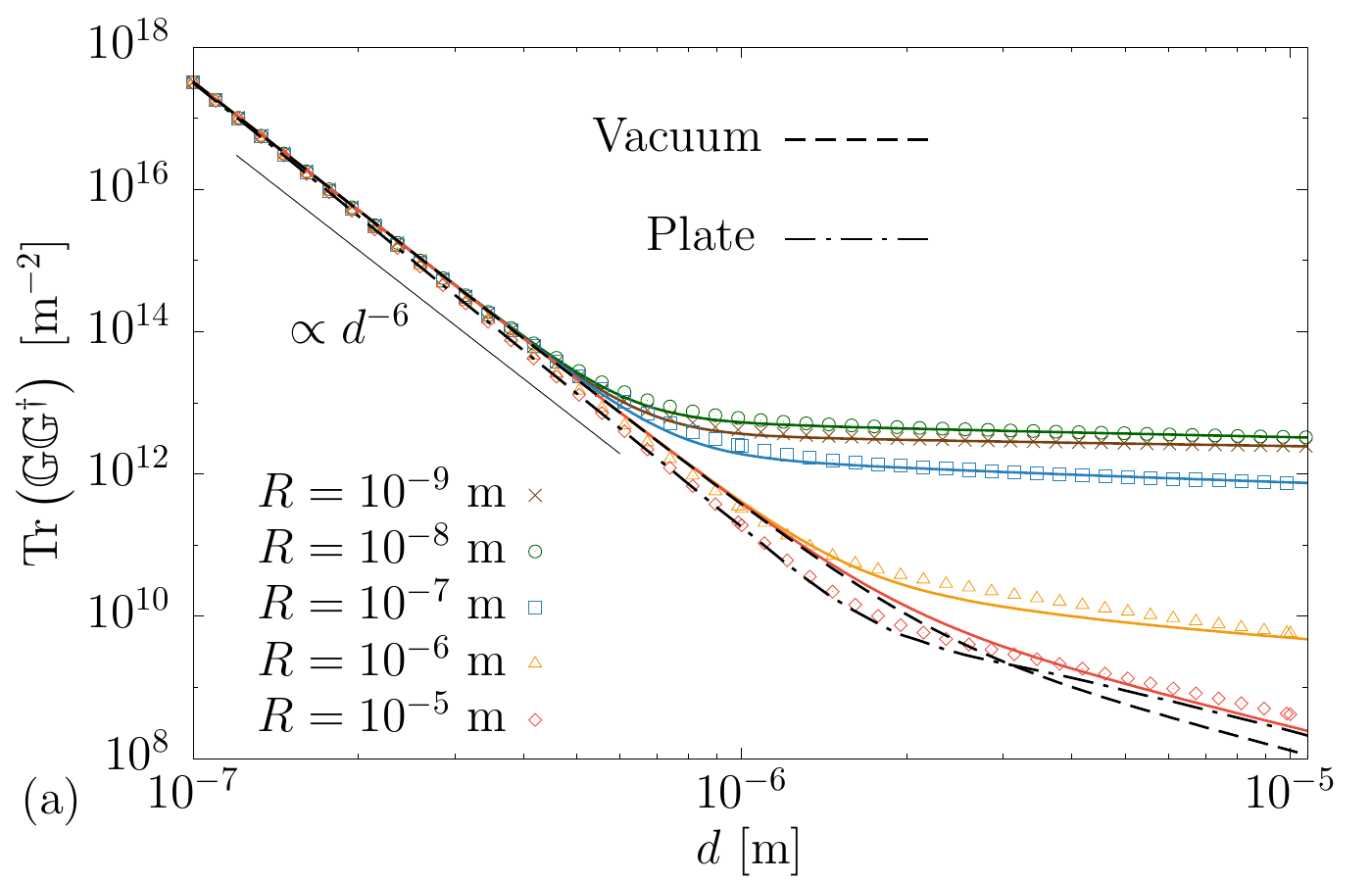}
\includegraphics[width=0.497\linewidth]{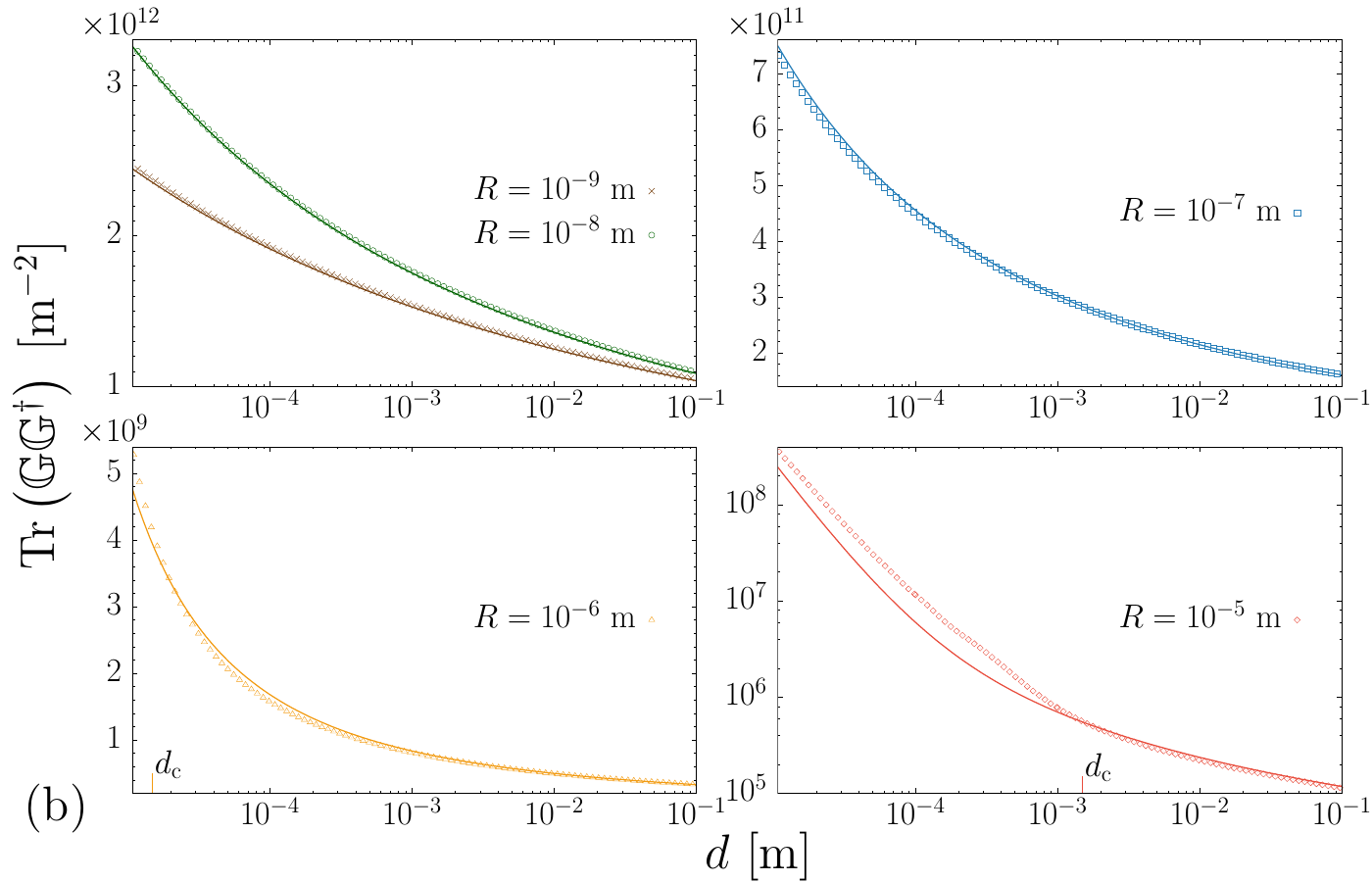}
\end{center}
\caption{\label{fig:GF_ddep_nfff}A more detailed version of Fig.~\ref{fig:GF_ddep}. Panel (a) shows the near-field results ($ d \leq \lambda_0 $), while panel (b) gives the far-field dependence ($ d \geq \lambda_0 $). The marks on horizontal axes correspond to $ d_{\textrm{c}} $ given by Eq.~\eqref{eq:dc}.}
\end{figure}

Figure~\ref{fig:GF_ddep} shows the dependence of $ \Tr\left(\mathbb{G}\mathbb{G}^{\dagger}\right) $ on $ d $ for the fixed $ h = 10^{-7} \ \textrm{m} $ and different $ R $, which is very similar to the behavior of the HT in Fig.~\ref{fig:HT_ddep} in the main text. Figure~\ref{fig:GF_ddep_nfff} provides with a different representation of Fig.~\ref{fig:GF_ddep}, separating the near- and far-field results, as well as giving some plots with logarithmic horizontal axis and linear vertical axis.  This makes the similarity and the distinction between the numerical and analytical curves more visible.

\begin{figure}[!b]
\begin{center}
\includegraphics[width=0.497\linewidth]{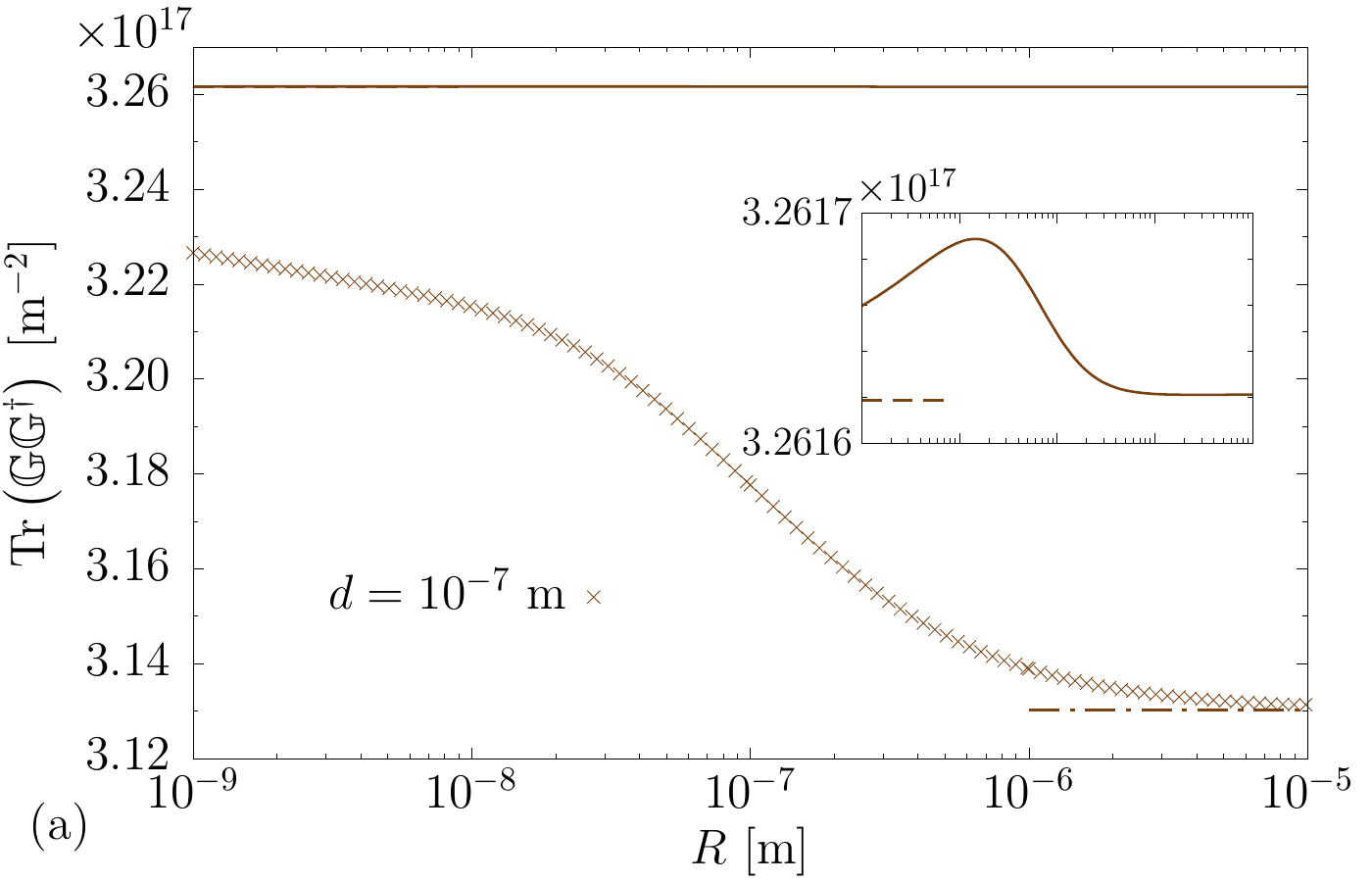}
\includegraphics[width=0.497\linewidth]{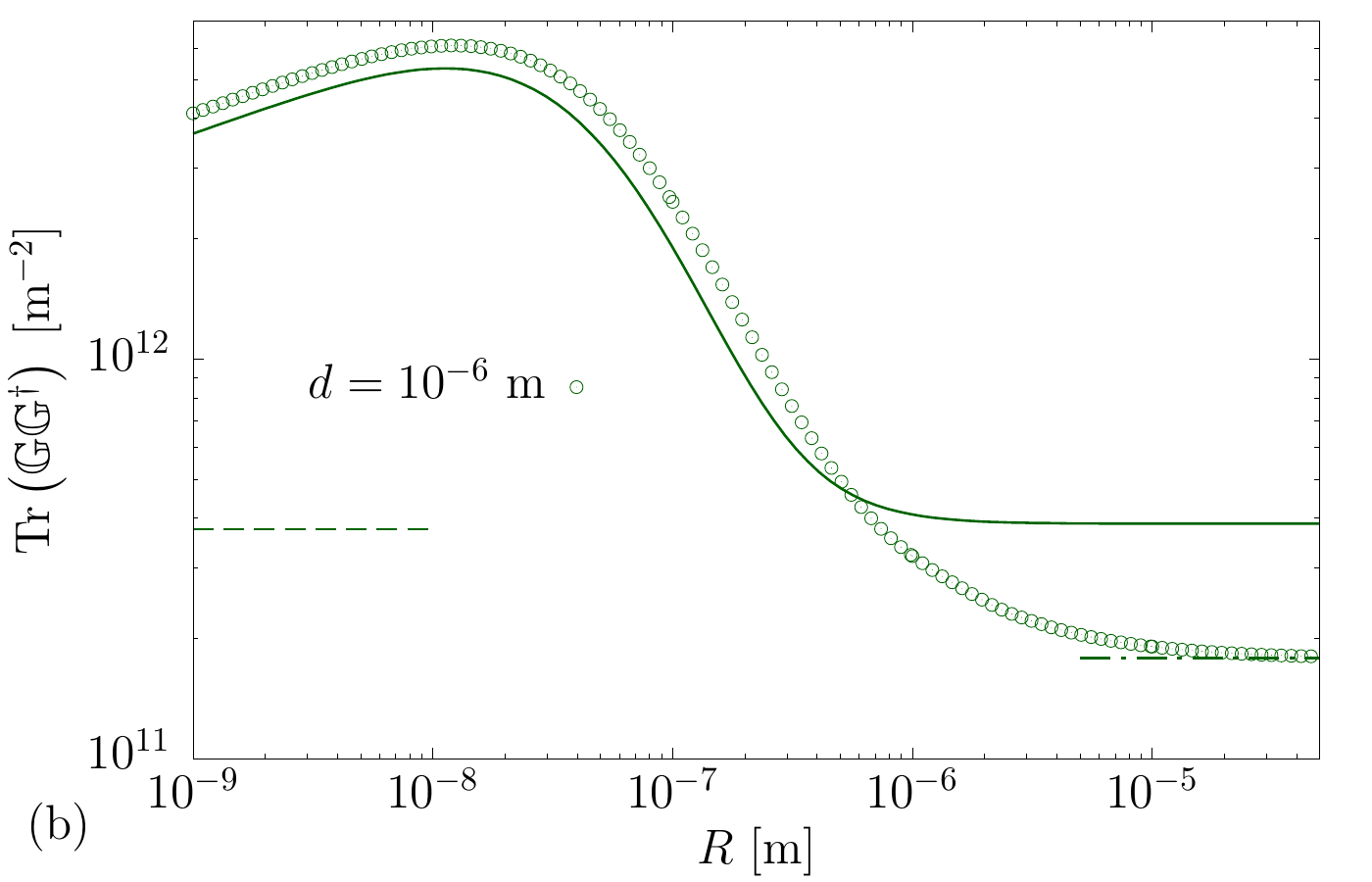}\\
\vspace{7pt}
\includegraphics[width=0.329\linewidth]{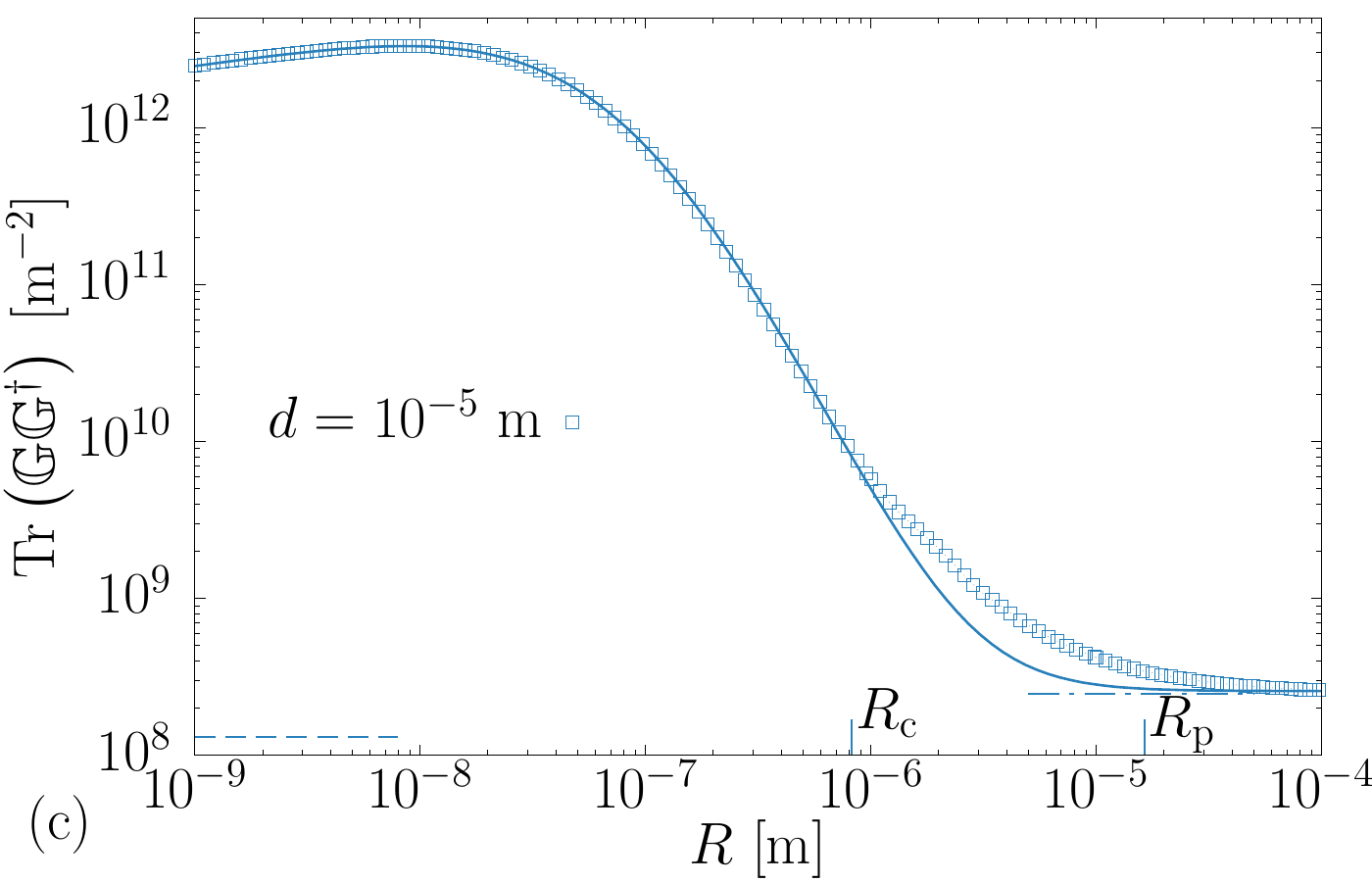}
\includegraphics[width=0.329\linewidth]{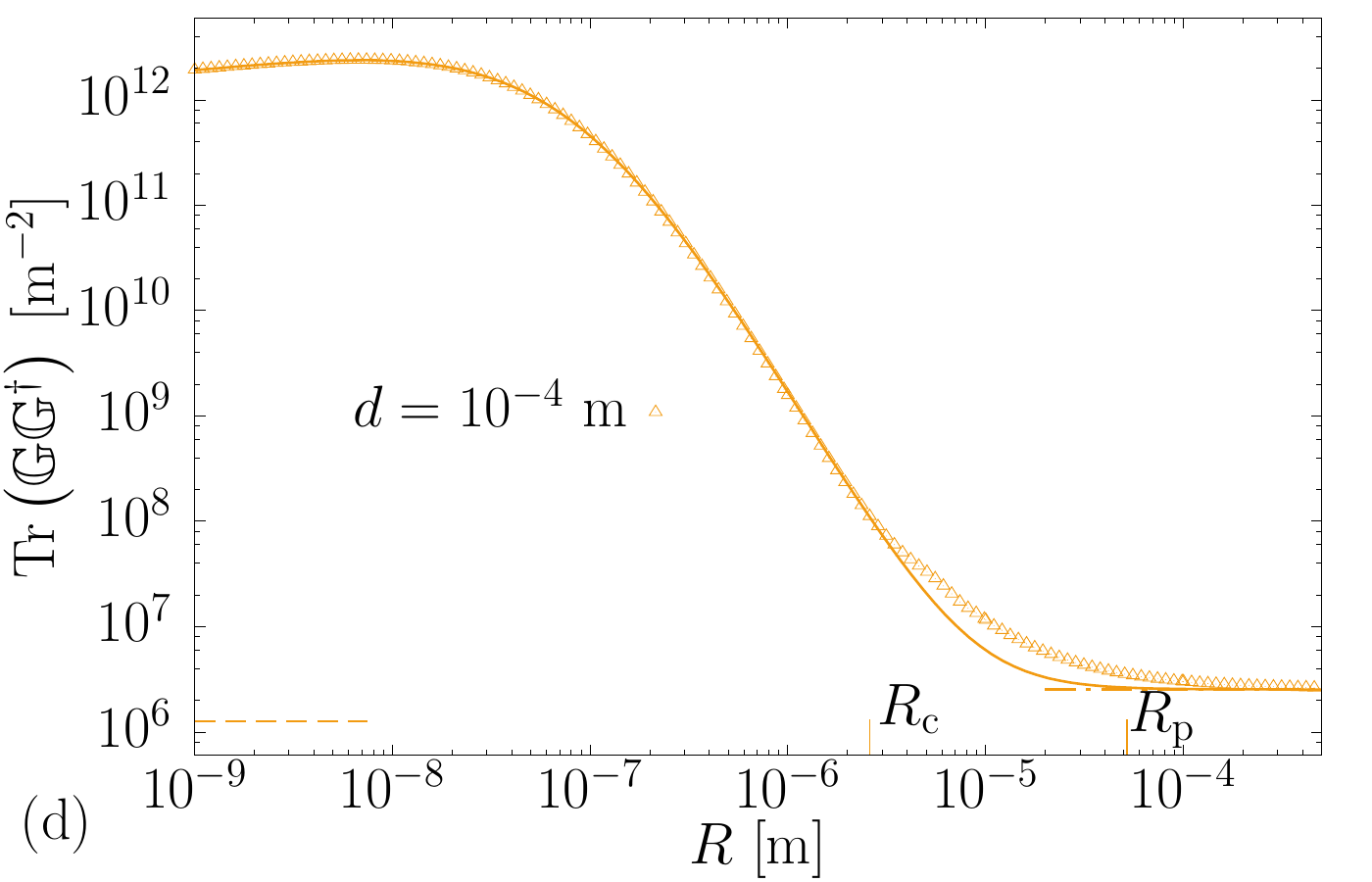}
\includegraphics[width=0.329\linewidth]{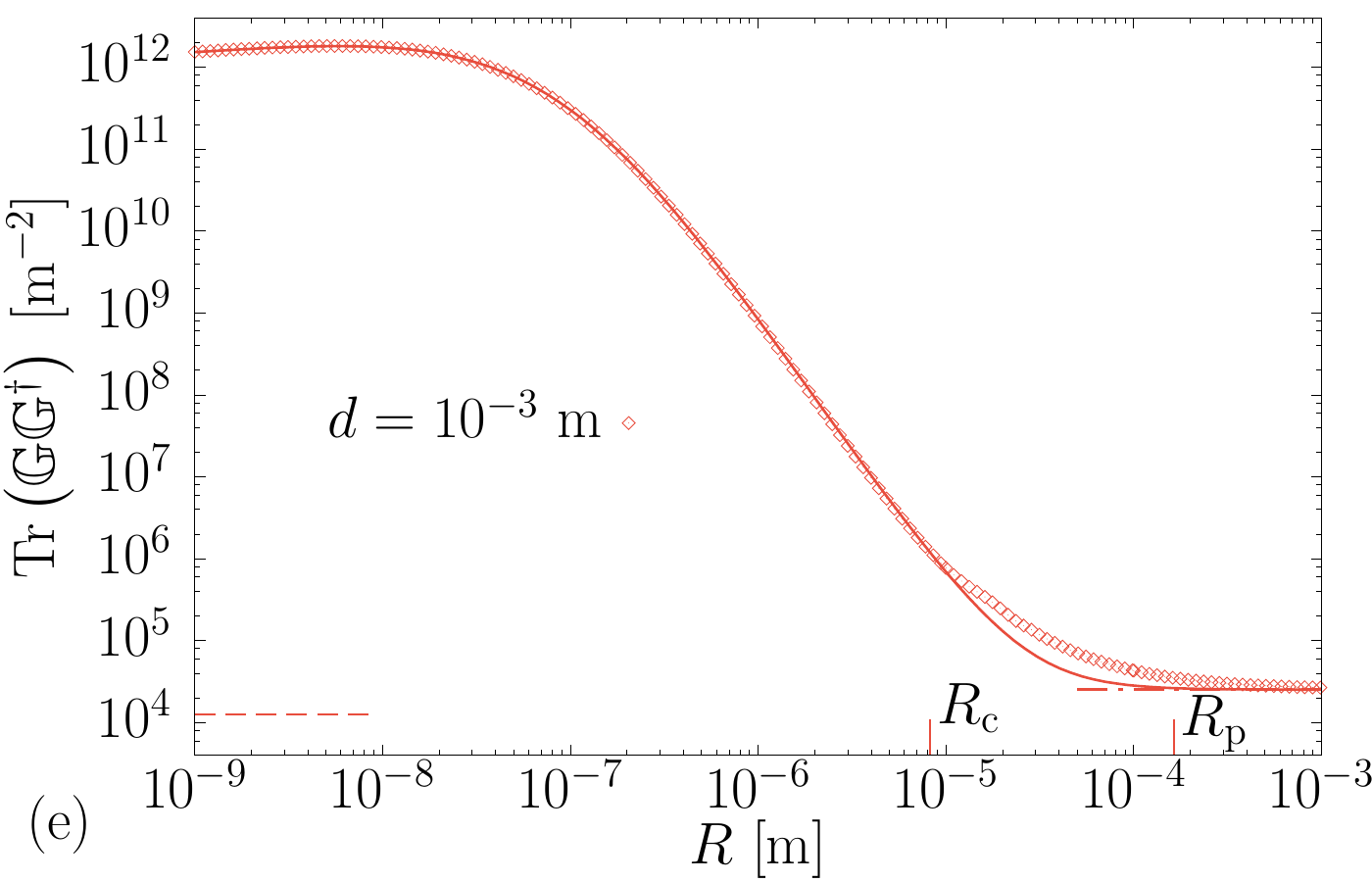}
\end{center}
\caption{\label{fig:GF_Rdep}$ \Tr\left(\mathbb{G}\mathbb{G}^{\dagger}\right) $ of a perfectly conducting cylinder as a function of its radius $ R $. The results are given for different distances $ d $ between the particles (near-field $ d $ in top figures and far-field $ d $ in bottom figures). Points show numerically exact results, while solid lines represent analytical approximation given by Eq.~\eqref{eq:Tr} in the main text. Dashed and dashed-dotted lines show the vacuum and plate results, respectively. The marks on horizontal axes correspond to $ R_{\textrm{c}} $ and $ R_{\textrm{p}} $, given by Eqs.~\eqref{eq:Rc} and~\eqref{eq:Rp}, respectively. The inset in panel (a) shows a detailed version of the corresponding analytical approximation. Other parameters are the same as in Fig.~\ref{fig:GF_ddep}.}
\end{figure}

In Fig.~\ref{fig:GF_Rdep}, the dependence on $ R $ for different $ d $ is shown. As it can be concluded also from Fig.~\ref{fig:GF_ddep}, the relative variation with $ R $ becomes larger with increase of $ d $. For large $ R $, the convergence to the case of PPs above a plate can be observed. The vacuum result is expected when $ R $ is small, but far below $ R = 10^{-9} \ \textrm{m} $ (as can be seen in Fig.~\ref{fig:GF_Rdep}), which can be regarded as physically inaccessible radii. When $ d = 10^{-7} \ \textrm{m} $, the trace decays monotonically to the plate result, while a maximum appears when $ d = 10^{-6} \ \textrm{m} $. The behavior of $ \Tr\left(\mathbb{G}\mathbb{G}^{\dagger}\right) $ for far-field separations (bottom figures) can be compared to that of the HT in Fig.~\ref{fig:HT_Rdep} in the main text: The two quantities have very similar dependencies on $ R $.

\begin{figure}[!t]
\begin{center}
\includegraphics[width=0.497\linewidth]{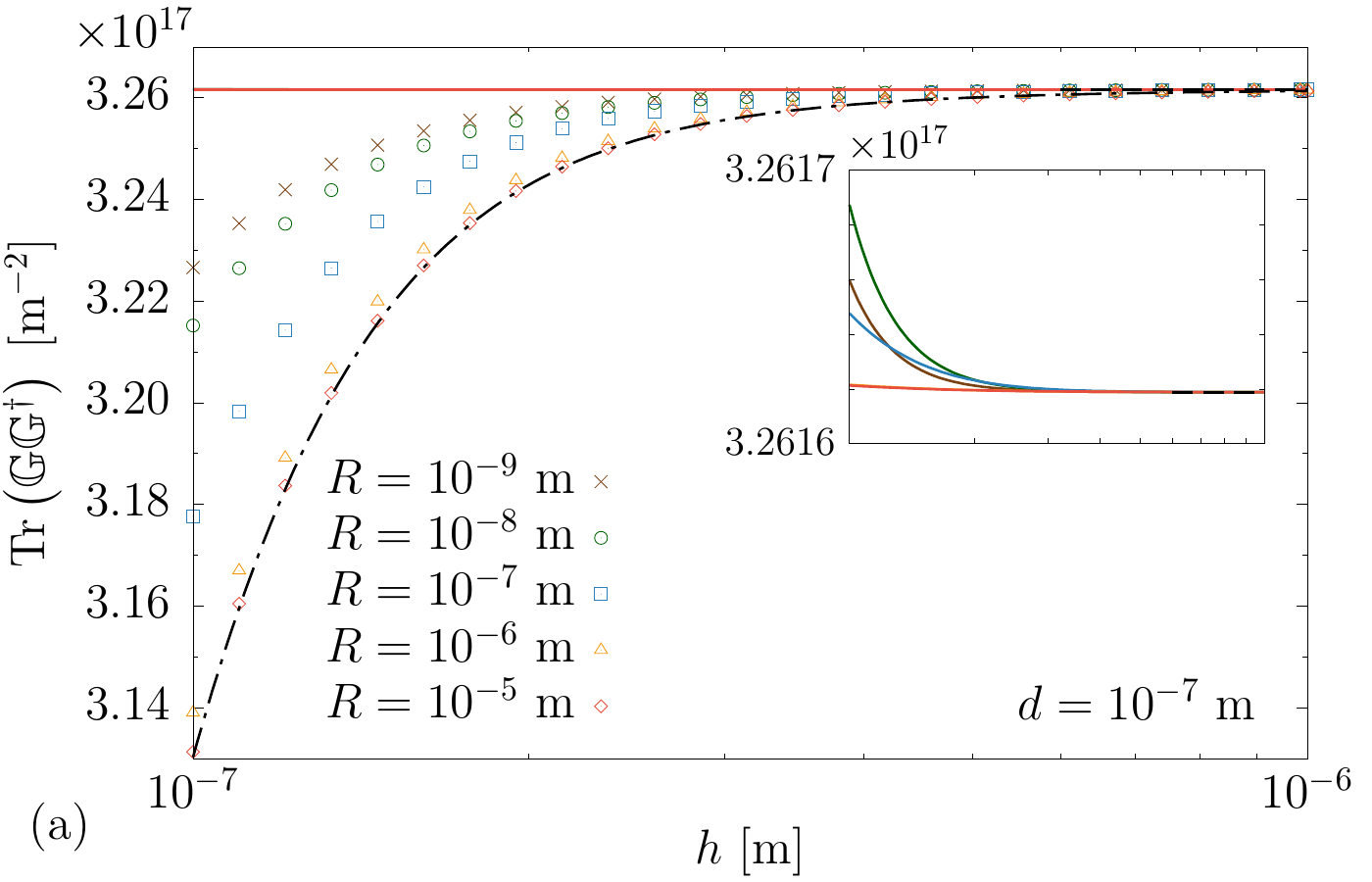}
\includegraphics[width=0.497\linewidth]{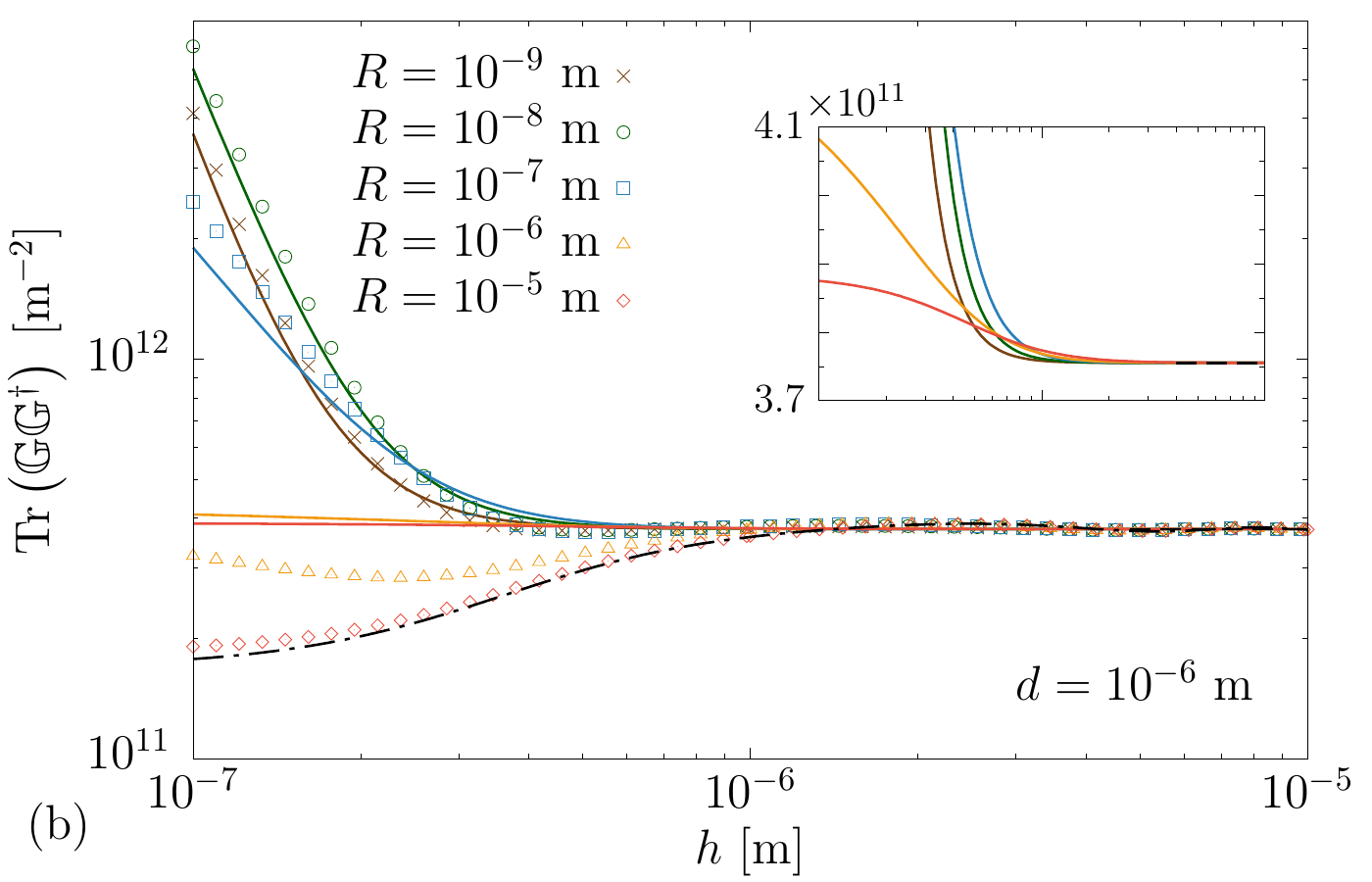}\\
\vspace{7pt}
\includegraphics[width=0.329\linewidth]{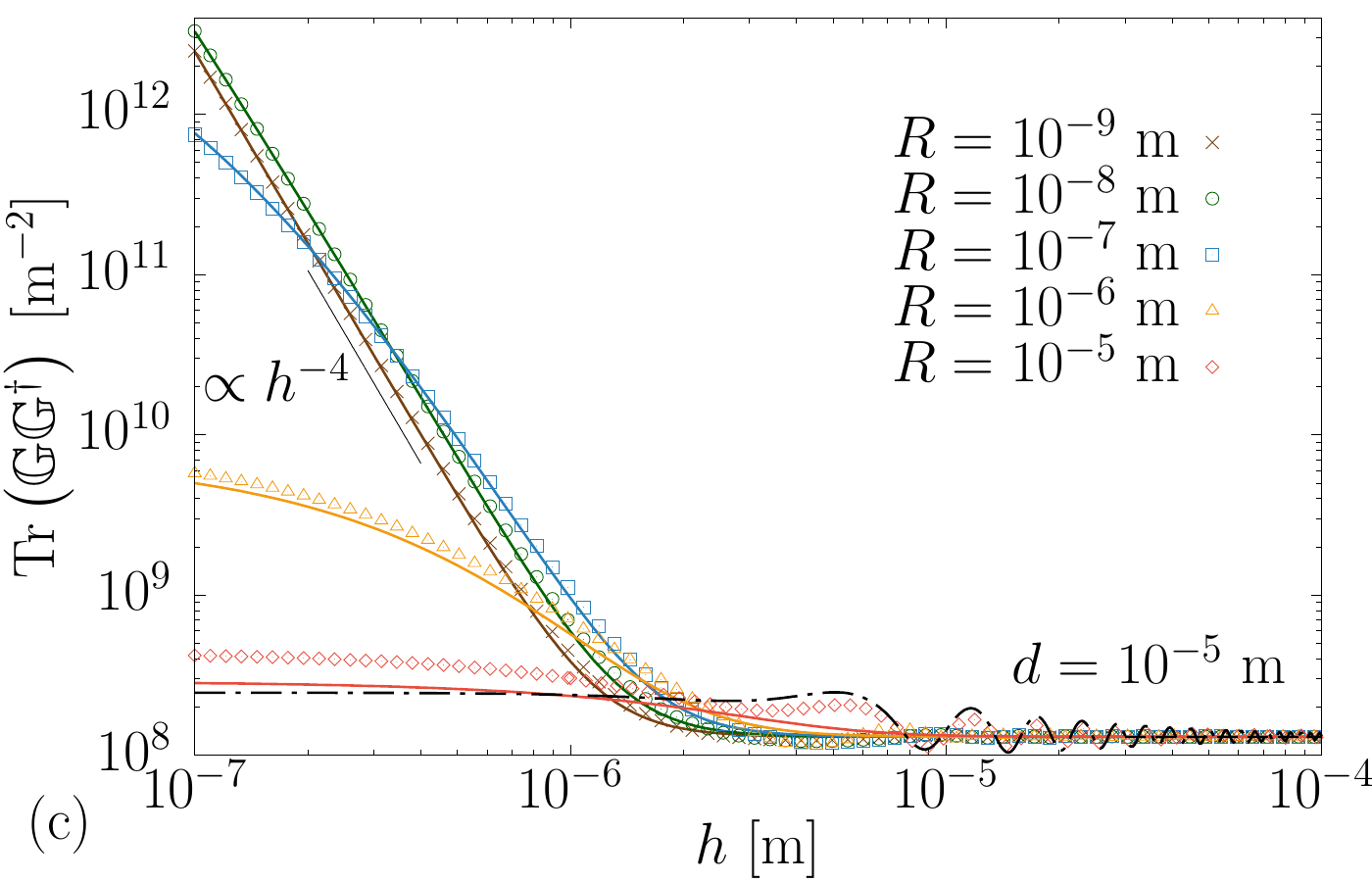}
\includegraphics[width=0.329\linewidth]{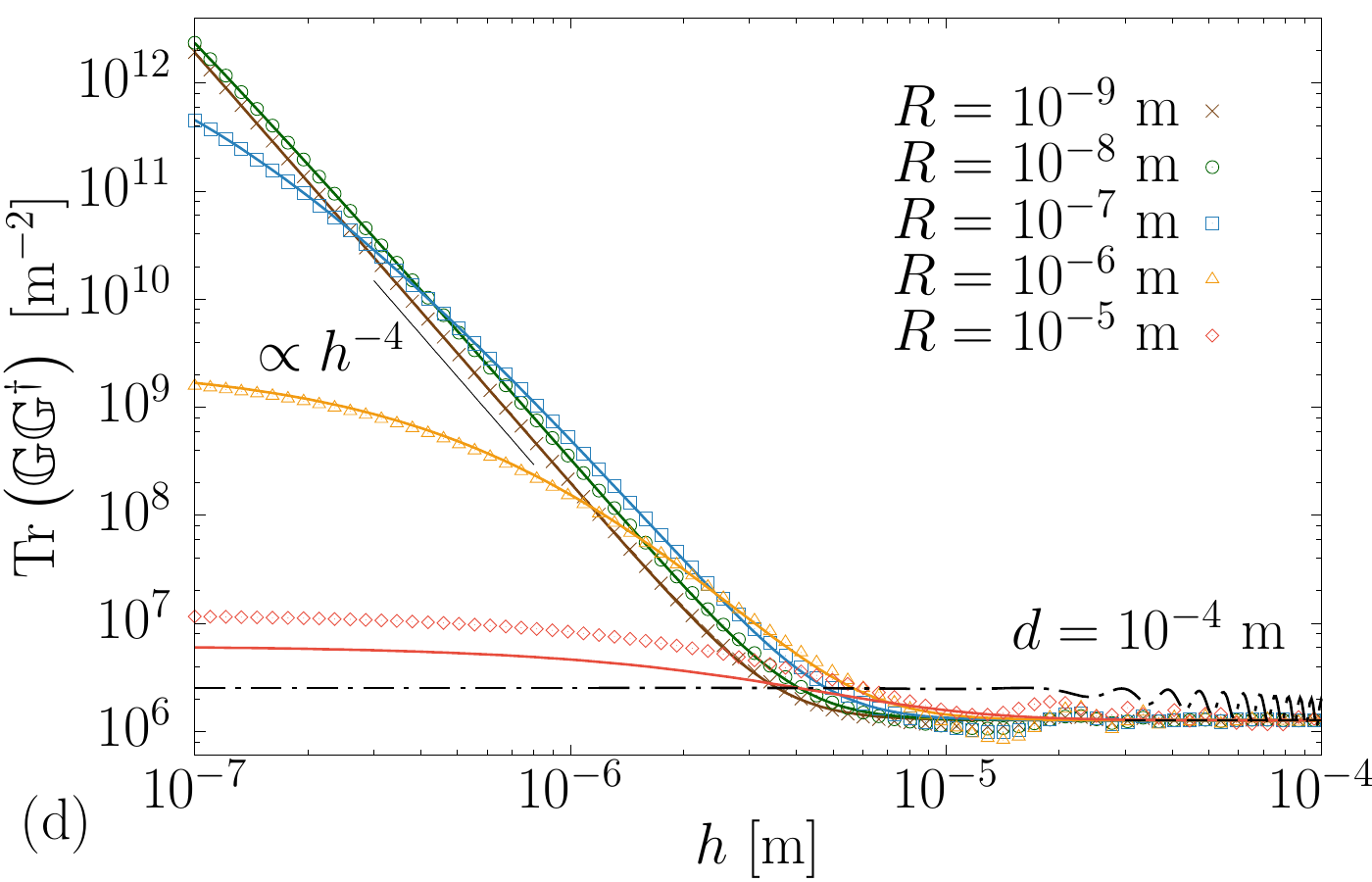}
\includegraphics[width=0.329\linewidth]{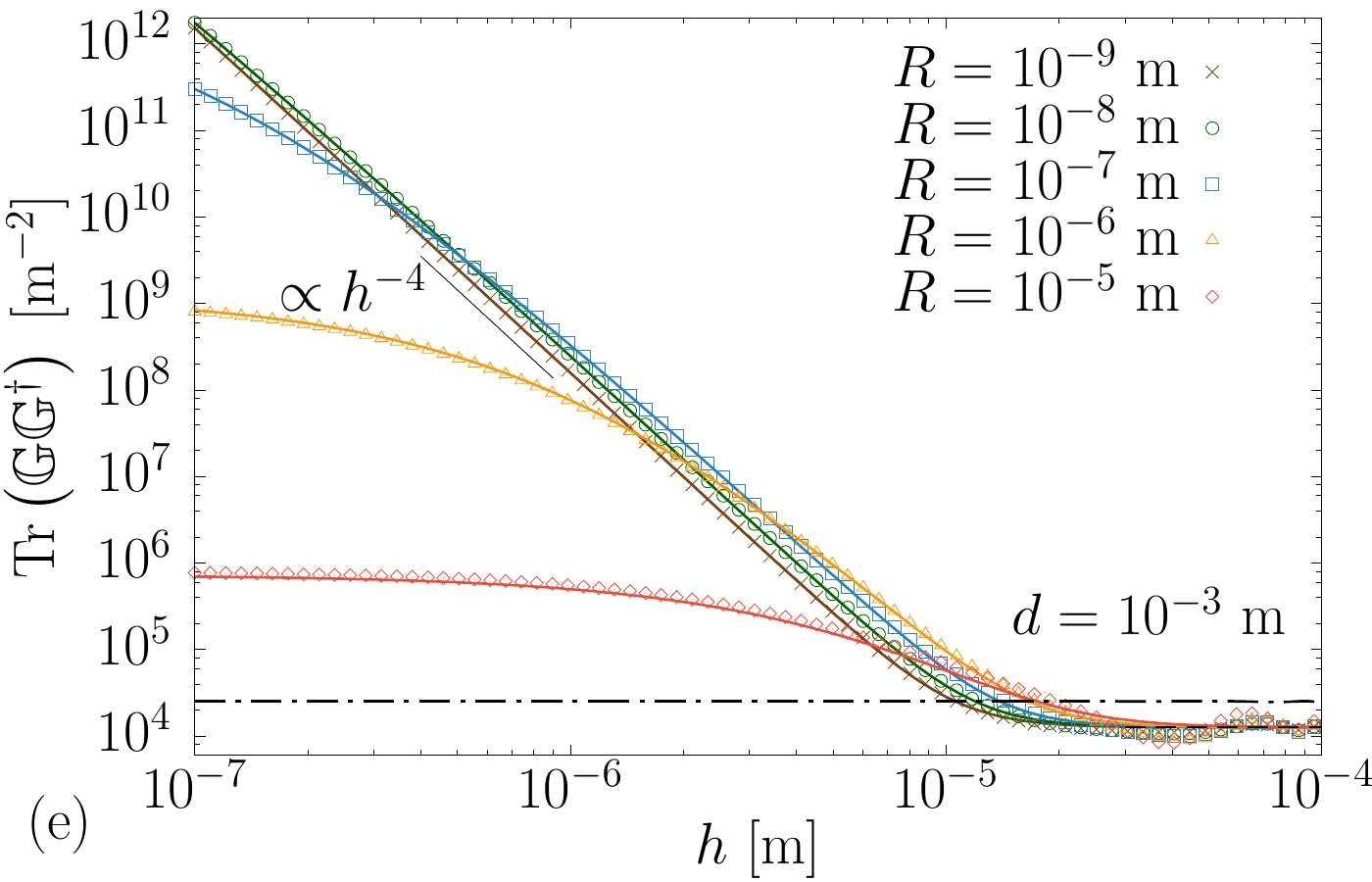}
\end{center}
\caption{\label{fig:GF_hdep}$ \Tr\left(\mathbb{G}\mathbb{G}^{\dagger}\right) $ of a perfectly conducting cylinder as a function of distance $ h $ from the particles to its surface. The results are given for different distances $ d $ between the particles (near-field $ d $ in top figures and far-field $ d $ in bottom figures) and radii $ R $ of the cylinder. Points show numerically exact results, while solid lines represent analytical approximation given by Eq.~\eqref{eq:Tr} in the main text. Dashed and dashed-dotted lines show the vacuum and plate results, respectively. The insets in panels (a) and (b) contain details of the analytical curves. Other parameters are the same as in Fig.~\ref{fig:GF_ddep}.}
\end{figure}

The dependence on $ h $ is given in Fig.~\ref{fig:GF_hdep}. For $ d = 10^{-7} \ \textrm{m} $, $ \Tr\left(\mathbb{G}\mathbb{G}^{\dagger}\right) $ converges fast to the vacuum result with increasing $ h $. Similarly to the $ R $ dependence in Fig.~\ref{fig:GF_Rdep}(a), the relative variation with $ h $ is small and the discrepancy between the numerical and analytical curves catches the eye. The picture changes when $ d = 10^{-6} \ \textrm{m} $, where a solid deviation from the vacuum result is observed for small $ h $: The trace is larger for small $ R $, but it is smaller for large $ R $. Also, small oscillations become visible. The discrepancy between the numerics and analytics remains significant. This discrepancy is much smaller in the far field. Here, the behavior is qualitatively the same for any particular $ d $, as shown in bottom panels of Fig.~\ref{fig:GF_hdep}. With increasing $ h $, a monotonic decay is followed by fading oscillations, finally converging to the vacuum result. The first minimum of the oscillations and the convergence are shifted to larger $ h $ if $ d $ is increased (roughly speaking, the overall dependence on $ h $ is shifted to the right with increase of $ d $). If there is the cylinder case (Eq.~\eqref{eq:cylinder_case} is satisfied) and  $ h \gg R $, the decay is of the form $h^{-4}$ (compare to Eq.~\eqref{eq:Tr_lambda_ff}). When $ h \ll R $, the trace goes to a constant with further decrease of $ h $. The plate curve is approached with increase of $ R $; this approach is faster when $ d $ is smaller (in agreement with Fig.~\ref{fig:GF_ddep}). We do not show $ \Tr\left(\mathbb{G}\mathbb{G}^{\dagger}\right) $ below $ h = 10^{-7} \ \textrm{m} $, because, in this case, PP limit would require very small sizes of the particles. However, formally, the results can be obtained for much smaller values of $ h $; for $ h \to 0 $, the convergence to a constant is expected (compare to Eq.~\eqref{eq:Tr_lambda_ff}).

\subsubsection{Analytical approximation}

\paragraph{General remarks\\}

Assuming that the slow decay in Fig.~\ref{fig:GF_ddep} has a logarithmic form, we found that $ \Tr\left(\mathbb{G}\mathbb{G}^{\dagger}\right) $ can be approximated by Eq.~\eqref{eq:Tr} in the main text. We note that, for any GF, the trace can be decomposed into three contributions~\cite{Messina2018}:
\begin{equation}
\Tr\big(\mathbb{G}\mathbb{G}^{\dagger}\big) =  \Tr\big(\mathbb{G}_0\mathbb{G}_0^{\dagger}\big) +  \Tr\big(\mathbb{G}_{\mathbb{T}}\mathbb{G}_{\mathbb{T}}^{\dagger}\big) + 2\Re\big[\Tr\big(\mathbb{G}_0\mathbb{G}_{\mathbb{T}}^{\dagger}\big)\big].
\label{eq:Tr_decompose}
\end{equation}
In Eq.~\eqref{eq:Tr} in the main text, the first (power-law) term is the exact result for $ \Tr\big(\mathbb{G}_0\mathbb{G}_0^{\dagger}\big) $. The second (logarithmic) term is the far-field approximation for $ \Tr\big(\mathbb{G}_{\mathbb{T}}\mathbb{G}_{\mathbb{T}}^{\dagger}\big) $. The contribution of $ 2\Re\big[\Tr\big(\mathbb{G}_0\mathbb{G}_{\mathbb{T}}^{\dagger}\big)\big] $ is missing in Eq.~\eqref{eq:Tr} in the main text, such that the full trace is larger than the vacuum part, leading to the corresponding property of the approximated HT. This is, however, not true in general for exact values of $ \Tr\left(\mathbb{G}\mathbb{G}^{\dagger}\right) $ and HT, as can be seen from the figures. Note that the near-field contribution of $ \Tr\big(\mathbb{G}_{\mathbb{T}}\mathbb{G}_{\mathbb{T}}^{\dagger}\big) $ is absent in Eq.~\eqref{eq:Tr} in the main text, such that the equation is expected to work worse for near-field $ d $, especially when one considers low frequencies (which is equivalent to using small $ T_1 $ in the HT). Indeed, we observed that formula~\eqref{eq:HTanapproxGF} in the main text approximates the HT worse if $ T_1 $ decreases. However, the approximation is still good even for $ T_1 $ which is $ 100 $ times smaller than room temperature.

Overall, Eq.~\eqref{eq:Tr} in the main text is a good approximation for the trace, as Figs.~\ref{fig:GF_ddep},~\ref{fig:GF_ddep_nfff},~\ref{fig:GF_Rdep}, and~\ref{fig:GF_hdep} show. Although strong deviations from numerical data appear for near-field interparticle distances, especially regarding $ R $ and $ h $ dependence (see top panels of Figs.~\ref{fig:GF_Rdep} and~\ref{fig:GF_hdep}), the relative discrepancy is not large. For $ d \gtrsim \lambda $, formula~\eqref{eq:Tr} in the main text approximates $ \Tr\left(\mathbb{G}\mathbb{G}^{\dagger}\right) $ very good, especially within certain ranges of the parameters, as discussed below.\\

\paragraph{Three cases for the far field\\}

The condition $ d \gtrsim \lambda $ allows to simplify and analyze Eq.~\eqref{eq:Tr} in the main text. Rewriting the formula in terms of the wavelength,
\begin{equation}
\Tr\left(\mathbb{G}\mathbb{G}^{\dagger}\right) \approx \frac{1}{8\pi^2 d^2}\left[1+\frac{\lambda^2}{4\pi^2 d^2}+\frac{3\lambda^4}{16\pi^4 d^4}\right] + \frac{\lambda^2}{16\pi^4(R+h)^4\ln^2\left[1+\frac{\lambda\sqrt{d^2+4h^2}}{\sqrt{2}\pi R^2}\right]},
\label{eq:Tr_lambda}
\end{equation}
and taking the far-field limit ($ d \gtrsim \lambda $), we get
\begin{equation}
\Tr\left(\mathbb{G}\mathbb{G}^{\dagger}\right) \approx \frac{1}{8\pi^2 d^2} + \frac{\lambda^2}{16\pi^4(R+h)^4\ln^2\left[1+\frac{\lambda\sqrt{d^2+4h^2}}{\sqrt{2}\pi R^2}\right]}.
\label{eq:Tr_lambda_ff}
\end{equation}
On the other hand, for the far-field-separated particles above a plate, the trace can be approximated as
\begin{equation}
\Tr\big(\mathbb{G}_{\textrm{pl}}\mathbb{G}_{\textrm{pl}}^{\dagger}\big) \approx \frac{1}{8\pi^2 d^2} + \frac{1}{8\pi^2\left(d^2+4h^2\right)},
\label{eq:Tr_plate_ff}
\end{equation}
where, as in the cylinder approximation~\eqref{eq:Tr_lambda_ff}, the first term is the vacuum part, while the second term is the approximation for the scattering part.

It is insightful to introduce three cases, based on Eqs.~\eqref{eq:Tr_lambda_ff} and~\eqref{eq:Tr_plate_ff}, and on the plots. The first case is when $ \Tr\left(\mathbb{G}\mathbb{G}^{\dagger}\right) \gg 2\Tr\big(\mathbb{G}_0\mathbb{G}_0^{\dagger}\big) \geq \Tr\big(\mathbb{G}_{\textrm{pl}}\mathbb{G}_{\textrm{pl}}^{\dagger}\big) $, such that the logarithmic term in Eq.~\eqref{eq:Tr_lambda_ff} dominates. This corresponds to a large enhancement of the trace compared to isolated particles and to a slow stable decay with $ d $, observed for large $ d $, small $ R $, and small $ h $ (see Figs.~\ref{fig:GF_ddep},~\ref{fig:GF_Rdep}, and~\ref{fig:GF_hdep}), which can be regarded as the features of a cylindrical waveguide. Therefore, we call this case~\enquote{cylinder case}. A more precise condition for this situation can be written as
\begin{equation}
\ln\left[1+\frac{\lambda\sqrt{d^2+4h^2}}{\sqrt{2}\pi R^2}\right] \ll \frac{\lambda d}{2\pi \left(R+h\right)^2}.
\label{eq:cylinder_case}
\end{equation}
For an important scenario where $ h \ll \lambda $, considering typical $ \lambda $ (around $ \lambda_{T_1} $) and physical $ R $, Eq.~\eqref{eq:cylinder_case} can be approximately simplified to
\begin{equation}
\lambda d \gg 16R^2.
\label{eq:cylinder_case_nfh}
\end{equation}
From Eq.~\eqref{eq:cylinder_case_nfh}, it follows that, for a fixed $ R $, the cylinder case is given when $ d \gtrsim d_{\textrm{c}} $, where
\begin{equation}
d_{\textrm{c}} = \frac{160R^2}{\lambda}.
\label{eq:dc}
\end{equation}
This quantity is marked on horizontal axes of Fig.~\ref{fig:GF_ddep_nfff}(b). Notably, approximation~\eqref{eq:Tr} in the main text works very good when $ d \gtrsim d_{\textrm{c}} $, but not as good for smaller $ d $. Overall, the agreement is expected to be better when $ d \notin [d_{\textrm{p}}, d_{\textrm{c}}] $, where $ d_{\textrm{p}} $ is given in Eq.~\eqref{eq:dp}, than in case $ d \in [d_{\textrm{p}}, d_{\textrm{c}}] $. Similarly, for a fixed $ d $, we have the cylinder case if $ R \lesssim R_{\textrm{c}} $, where
\begin{equation}
R_{\textrm{c}} = \frac{\sqrt{\lambda d}}{4\sqrt{10}},
\label{eq:Rc}
\end{equation}
marked on horizontal axes of Fig.~\ref{fig:GF_Rdep} (bottom plots). For $ R \lesssim R_{\textrm{c}} $, the trace is much larger than the plate trace, and analytical curves are in an excellent agreement with numerical data.

For large $ R $, a cylinder should give the same effect as a plate does. Indeed, taking the limit $ R \to \infty $ in Eq.~\eqref{eq:Tr_lambda_ff} leads to the plate trace in Eq.~\eqref{eq:Tr_plate_ff}. More precisely, our second case,~\enquote{plate case}, takes place when
\begin{equation}
\ln\left[1+\frac{\lambda\sqrt{d^2+4h^2}}{\sqrt{2}\pi R^2}\right] \approx \frac{\lambda\sqrt{d^2+4h^2}}{\sqrt{2}\pi R^2}, \ \ \ \ R \gg h.
\label{eq:plate_case}
\end{equation}
Note that equality $ \Tr\left(\mathbb{G}\mathbb{G}^{\dagger}\right) \approx \Tr\big(\mathbb{G}_{\textrm{pl}}\mathbb{G}_{\textrm{pl}}^{\dagger}\big) $ is necessary but not sufficient: It can be achieved also for small $ R $, which, however, does not correspond to a plate. Therefore, condition $ R \gg h $ is required in Eq.~\eqref{eq:plate_case}. If, furthermore, $ h \ll \lambda $, Eq.~\eqref{eq:plate_case} simplifies to
\begin{equation}
\lambda d \ll 4R^2,
\label{eq:plate_case_nfh}
\end{equation}
which, for a given $ R $, works if $ d \lesssim d_{\textrm{p}} $, where
\begin{equation}
d_{\textrm{p}} = \frac{2R^2}{5\lambda}.
\label{eq:dp}
\end{equation}
For a fixed $ d $, the plate case corresponds to $ R \gtrsim R_{\textrm{p}} $, where
\begin{equation}
R_{\textrm{p}} = \frac{\sqrt{10}}{2}\sqrt{\lambda d},
\label{eq:Rp}
\end{equation}
marked on horizontal axes of Fig.~\ref{fig:GF_Rdep} (bottom plots). One can see that, for $ R \gtrsim R_{\textrm{p}} $, the cylinder curves bend over the plate results, and the agreement with the analytical approximation becomes better. Note that, overall, the agreement is better when $ R \notin [R_{\textrm{c}}, R_{\textrm{p}}] $.

The third case is the vacuum case, where the logarithmic term in~\eqref{eq:Tr_lambda_ff} is much smaller than the quadratic one, or equivalently,
\begin{equation}
\ln\left[1+\frac{\lambda\sqrt{d^2+4h^2}}{\sqrt{2}\pi R^2}\right] \gg \frac{\lambda d}{\sqrt{2}\pi \left(R+h\right)^2}.
\label{eq:vacuum_case}
\end{equation}
It is achieved at large $ h $, as can be seen in Fig.~\ref{fig:GF_hdep}. When $ h \ll \lambda $, the vacuum case cannot be observed for physical values of $ R $ (the radius has to be extremely small in order to satisfy condition~\eqref{eq:vacuum_case}).\\

\paragraph{The far field with cylinder versus the near field in vacuum\\}
Figure~\ref{fig:GF_ddep} suggests that $ \Tr\left(\mathbb{G}\mathbb{G}^{\dagger}\right) $ for far-field $ d $ can be comparable to or even larger than $ \Tr\left(\mathbb{G}_0\mathbb{G}_0^{\dagger}\right) $ for near-field $ d $. With Eq.~\eqref{eq:Tr_lambda}, this situation can be analyzed analytically. For near-field distance in vacuum $ d_{\textrm{nf}} $ (i.e., $ d_{\textrm{nf}} \ll \lambda $), we have 
\begin{equation}
\Tr\left(\mathbb{G}_0\mathbb{G}_0^{\dagger}\right) = \frac{3\lambda^4}{128\pi^6d_{\textrm{nf}}^6}.
\label{eq:Tr_vacuum_nf}
\end{equation}
On the other hand, for far-field distance with cylinder $ d $ (i.e., $ d \gtrsim \lambda $), and considering the cylinder case with $ h \ll \lambda $, we can write
\begin{equation}
\Tr\left(\mathbb{G}\mathbb{G}^{\dagger}\right) \approx \frac{\lambda^2}{16\pi^4(R+h)^4\ln^2\left[1+\frac{\lambda d}{\sqrt{2}\pi R^2}\right]}.
\label{eq:Tr_ff}
\end{equation}
We are interested in the situation where $ \Tr\left(\mathbb{G}\mathbb{G}^{\dagger}\right) \gtrsim \Tr\left(\mathbb{G}_0\mathbb{G}_0^{\dagger}\right) $. Using Eqs.~\eqref{eq:Tr_vacuum_nf} and~\eqref{eq:Tr_ff}, this condition can be written as
\begin{equation}
\frac{3}{8\pi^2}\ln^2\left[1+\frac{\lambda d}{\sqrt{2}\pi R^2}\right]\lambda^2(R+h)^4 \lesssim d_{\textrm{nf}}^6.
\label{eq:Tr_compare}
\end{equation}
In the cylinder case, $ \lambda d \gg 16R^2 $ (see Eq.~\eqref{eq:cylinder_case_nfh}), and, for relevant parameters, the term $ \frac{3}{8\pi^2}\ln^2\left[1+\frac{\lambda d}{\sqrt{2}\pi R^2}\right] $ is of the order ranging from $ 1 $ to $ 10^2 $. Therefore, Eq.~\eqref{eq:Tr_compare} can be approximately written in a simple from:
\begin{equation}
\lambda^2(R+h)^4 \ll d_{\textrm{nf}}^6.
\label{eq:Tr_compare_simple}
\end{equation}
Importantly, if Eq.~\eqref{eq:Tr_compare_simple} is valid, conditions $ h \ll \lambda $ and $ \lambda d \gg 16R^2 $ are satisfied \textit{a posteriori}. Note that $ d $ is absent in condition~\eqref{eq:Tr_compare_simple}, because the logarithm in Eq.~\eqref{eq:Tr_compare} is barely sensitive to the change of $ d $. Also note that $ \{R, h\} \ll d_{\textrm{nf}} $ is necessary in order for Eq.~\eqref{eq:Tr_compare_simple} to be valid. Similar discussions lead to Eq.~\eqref{eq:HT_dzoom} in the main text.

\subsection{Comparison between \texorpdfstring{Eqs.~\eqref{eq:HTanapproxGF} and~\eqref{eq:HTanapprox} in the main text}{Eqs. (5) and (6) in the main text}}
\begin{figure}[!htbp]
\begin{center}
\includegraphics[width=0.7\linewidth]{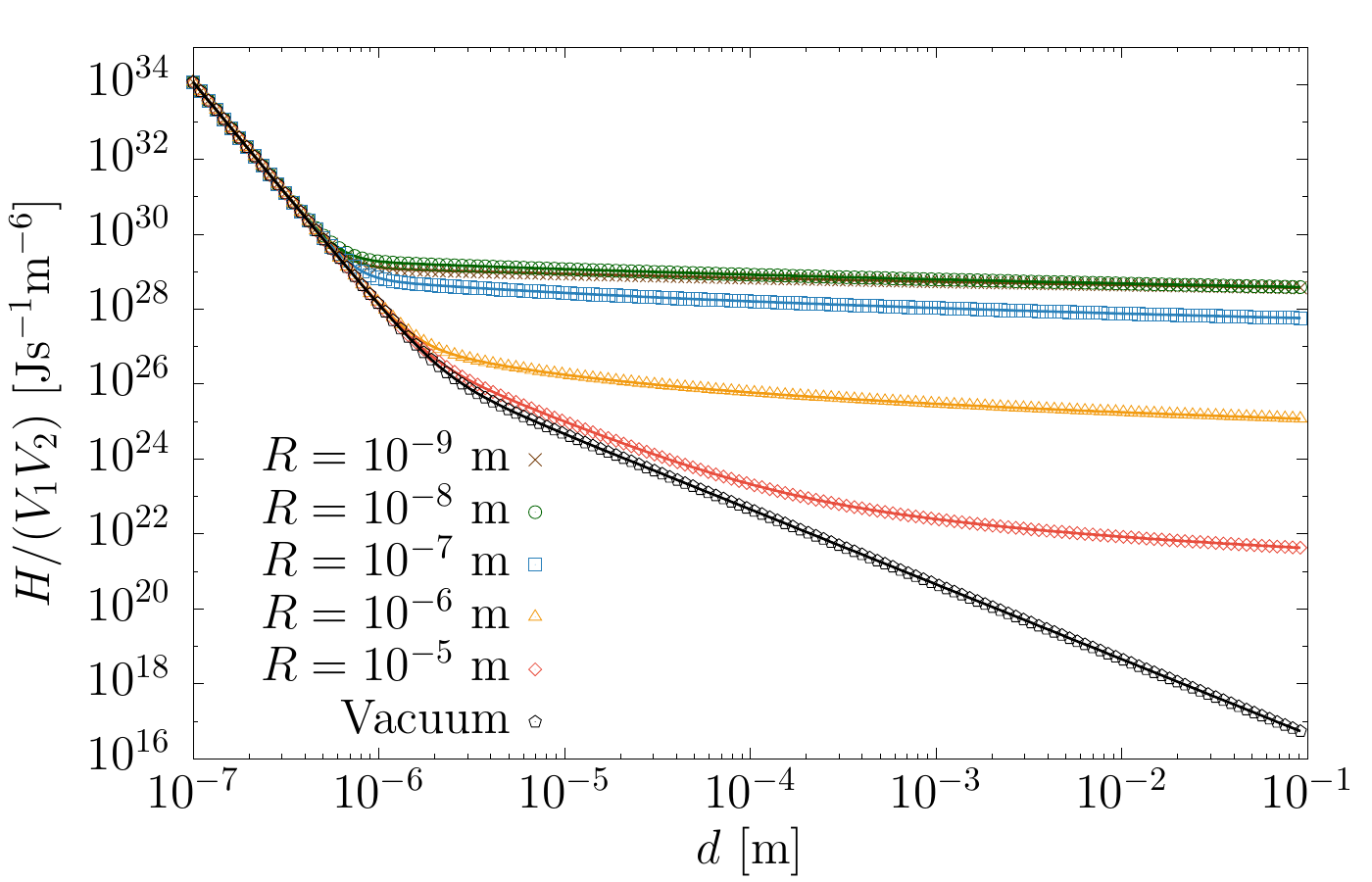}
\end{center}
\caption{\label{fig:Anapprox}Comparison between the heat transfer computed using Eqs.~\eqref{eq:HTanapproxGF} (points) and~\eqref{eq:HTanapprox} (lines) in the main text. The system and parameters are the same as in Fig.~\ref{fig:HT_ddep} in the main text. The vacuum results are obtained by neglecting the logarithmic terms in the equations.}
\end{figure}

Figure~\ref{fig:Anapprox} compares the HT computed with Eqs.~\eqref{eq:HTanapproxGF} and~\eqref{eq:HTanapprox} in the main text. The two equations show a perfect agreement, confirming that the latter equation is an excellent approximation for the former one.

\subsection{Comparison between the amounts of the transferred and radiated energy}
The energy transferred from SiC PP 1 at $ T_1 = 300 \ \textrm{K} $ to SiC PP 2 in the presence of a cylinder can be approximated by Eq.~\eqref{eq:HTanapprox} in the main text. On the other hand, the total energy emitted by PP 1 (taking into account the presence of a cylinder and PP 2) is given by~\cite{Asheichyk2017}
\begin{equation}
H_{\textrm{total}} = \frac{8\hbar}{c^2}\int_0^\infty d\omega \frac{\omega^3}{e^{\frac{\hbar\omega}{k_{\textrm{B}}T_1}}-1}\Im(\alpha_1)\Tr\Im\mathbb{G},
\label{eq:HR}
\end{equation}
where $ \mathbb{G} $ is the GF of a cylinder and PP 2. We consider $ d \gtrsim \lambda_0 $; it can be shown that, in this case, the contribution of PP 2 to $ \Tr\Im\mathbb{G} $, and hence to $ H_{\textrm{total}} $, is negligible. Therefore, we consider that $ \mathbb{G} $ is the GF of a cylinder (i.e., without PP 2). 

Similarly to the HT, the spectrum of $ H_{\textrm{total}} $ is strongly peaked at $ \omega_0 $, allowing us to approximate Eq.~\eqref{eq:HR} as
\begin{equation}
H_{\textrm{total}} \approx \frac{8\hbar}{c^2}\left[\Tr\Im\mathbb{G}(\omega_0)\right]\int_0^\infty d\omega \frac{\omega^3}{e^{\frac{\hbar\omega}{k_{\textrm{B}}T_1}}-1}\Im(\alpha_1).
\label{eq:HRtraceout}
\end{equation}
Therefore, the ratio between the HT and HR can be written as (since $ d \gtrsim \lambda_0 $, only the $ d^{-2} $ and logarithmic terms of the HT are considered)
\begin{equation}
\frac{H}{H_{\textrm{total}}} \approx \frac{R_2^3}{2\pi c^2}\frac{\left\{\frac{1}{d^2} + \frac{2c^2}{\omega_0^2(R+h)^4\ln^2\left[1+\frac{\sqrt{2}c\sqrt{d^2+4h^2}}{\omega_0R^2}\right]}\right\}}{\Tr\Im\mathbb{G}(\omega_0)}\frac{\int_0^\infty d\omega \frac{\omega^5}{e^{\frac{\hbar\omega}{k_{\textrm{B}}T_1}}-1}\left[\Im\left(\frac{\varepsilon_{\textrm{SiC}}-1}{\varepsilon_{\textrm{SiC}}+2}\right)\right]^2}{\int_0^\infty d\omega \frac{\omega^3}{e^{\frac{\hbar\omega}{k_{\textrm{B}}T_1}}-1}\Im\left(\frac{\varepsilon_{\textrm{SiC}}-1}{\varepsilon_{\textrm{SiC}}+2}\right)}.
\label{eq:HToverHR}
\end{equation}
In case of the particles above a plate, the logarithmic term in Eq.~\eqref{eq:HToverHR} is replaced by $ \frac{1}{d^2+4h^2} $ (according to Eq.~\eqref{eq:Tr_plate_ff}), while the exact analytical expression for $ \Tr\Im\mathbb{G} $ can be found in Ref.~\cite{Asheichyk2017}. For isolated particles, only $ \frac{1}{d^2} $ term in Eq.~\eqref{eq:HToverHR} remains, while $ \Tr\Im\mathbb{G}(\omega_0) = \frac{\omega_0}{2\pi c} $~\cite{Asheichyk2017}.

\begin{figure}[!t]
\begin{center}
\includegraphics[width=0.7\linewidth]{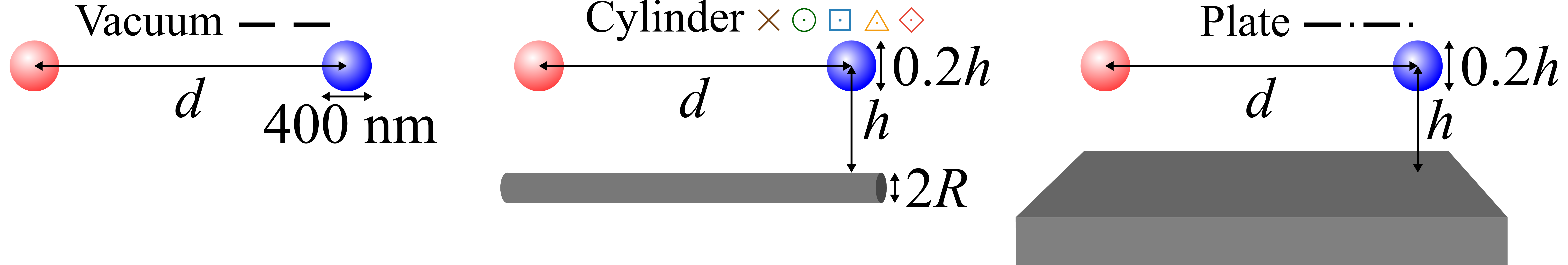}
\includegraphics[width=0.7\linewidth]{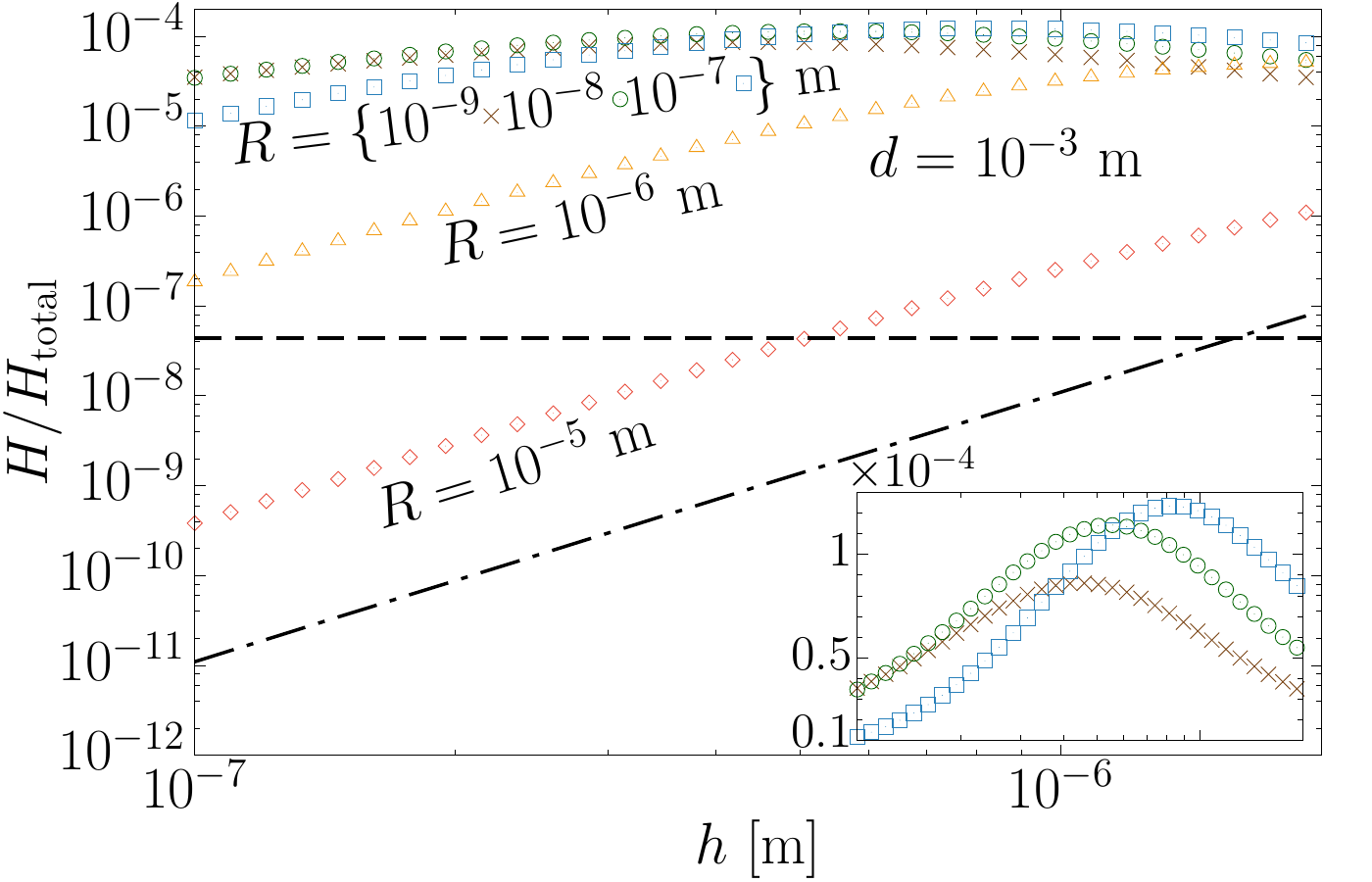}
\end{center}
\caption{\label{fig:HToverHR}The maximum ratio between the heat transfer and the total energy emitted by particle 1 in the presence of a perfectly conducting cylinder, as a function of $ h $, computed using Eq.~\eqref{eq:HToverHR}. Both particles are made of SiC; $ T_1 = 300 \ \textrm{K} $; $ d = 10^{-3} \ \textrm{m} $. The results are given for different radii of the cylinder and compared to cases of the particles in vacuum and above a perfectly conducting plate, with the same $ d $ (see the sketch). In cases of a cylinder or plate, $ R_2 = 0.1h $, while in the vacuum case, $ R_2 = 200 \ \textrm{nm} $. The inset shows zoomed version of the curves for small radii.}
\end{figure}

Note that the ratio in Eq.~\eqref{eq:HToverHR} is proportional to $ R_2^3 $, which is the feature of the PP limit. In that respect, the maximum of $ \frac{H}{H_{\textrm{total}}} $ is achieved when $ R_2 $ takes the maximum allowed value satisfying the PP condition. Since $ d \gtrsim \lambda_0 > \lambda_{T_1} > \delta_{\textrm{SiC}} $, where $ \delta_{\textrm{SiC}} = \frac{c}{\omega_0\Im\sqrt{\varepsilon(\omega_0)}} \approx 1.21 \times 10^{-6} \ \textrm{m} $ is the skin depth of SiC, this maximum value for a cylinder or a plate can be written as $ R_2 = 0.1h $, where $ h \lesssim \delta_{\textrm{SiC}} $ (for isolated particles, it can be set to $ R_2 = 200 \ \textrm{nm} $). These values are used in Fig.~\ref{fig:HToverHR}.

$ H/H_{\textrm{total}} $ is plotted in Fig.~\ref{fig:HToverHR} as a function of $ h $ for different radii $ R $ of the cylinder and $ d = 1 \ \textrm{mm} $. In order to estimate the maximum ratio, we take the largest possible $ R_2 $ satisfying the PP limit, as discussed above. As Fig.~\ref{fig:HToverHR} shows, the maximum ratio is the largest for thin cylinders, where it slightly exceeds $ 10^{-4} $. The vacuum and plate cases give much smaller values, which will decrease further as $ d^{-2} $ with increase of $ d $ (for a cylinder, this is prevented by the logarithmic dependence). It remains an open question, whether $ H/H_{\textrm{total}} $ for large $ d $ can be comparable to $ 1 $ beyond the PP limit. If yes, this would provide with an opportunity to drastically change the thermal state of a particle by placing another particle far away.

\subsection{Heat transfer in the presence of a cylinder of a nonperfectly conducting material}

\subsubsection{Green's function}
Let us consider configuration in Fig.~\ref{fig:System} in the main text and an infinitely long cylinder made of a homogeneous isotropic material with dielectric permittivity $ \varepsilon $ (we assume a nonmagnetic cylinder, i.e., magnetic permeability $ \mu = 1 $). In this case, Eqs.~\eqref{eq:G_separation},~\eqref{eq:GT_expansion}, and~\eqref{eq:GT_matrix} remain unchanged, whereas the scattering matrix elements~\eqref{eq:TMM},~\eqref{eq:TNN}, and~\eqref{eq:TMN} take a more complex form~\cite{Golyk2012, Noruzifar2011, Bohren2004}:
\begin{subequations}
\begin{alignat}{1}
& T_{n,k_z}^{MM} = -\frac{J_n(qR)}{H_n(qR)}\frac{\Delta_1\Delta_4-K^2}{\Delta_1\Delta_2-K^2},\label{eq:TMMe}\\
& T_{n,k_z}^{NN} = -\frac{J_n(qR)}{H_n(qR)}\frac{\Delta_2\Delta_3-K^2}{\Delta_1\Delta_2-K^2},\label{eq:TNNe}\\
& T_{n,k_z}^{MN} = T_{n,k_z}^{NM} = \frac{2i}{\pi\sqrt{\varepsilon}\left[qRH_n(qR)\right]^2}\frac{K}{\Delta_1\Delta_2-K^2},\label{eq:TMNe}
\end{alignat}
\end{subequations}
where
\begin{subequations}
\begin{alignat}{1}
& \Delta_1 = \frac{J'_n(q_{\varepsilon}R)}{q_{\varepsilon}RJ_n(q_{\varepsilon}R)}-\frac{1}{\varepsilon}\frac{H'_n(qR)}{qRH_n(qR)},\label{eq:Delta1}\\
& \Delta_2 = \frac{J'_n(q_{\varepsilon}R)}{q_{\varepsilon}RJ_n(q_{\varepsilon}R)}-\frac{H'_n(qR)}{qRH_n(qR)},\label{eq:Delta2}\\
& \Delta_3 = \frac{J'_n(q_{\varepsilon}R)}{q_{\varepsilon}RJ_n(q_{\varepsilon}R)}-\frac{1}{\varepsilon}\frac{J'_n(qR)}{qRJ_n(qR)},\label{eq:Delta3}\\
& \Delta_4 = \frac{J'_n(q_{\varepsilon}R)}{q_{\varepsilon}RJ_n(q_{\varepsilon}R)}-\frac{J'_n(qR)}{qRJ_n(qR)},\label{eq:Delta4}
\end{alignat}
\end{subequations}
and
\begin{equation}
K = \frac{nk_z}{\sqrt{\varepsilon}kR^2}\left(\frac{1}{q_{\varepsilon}^2}-\frac{1}{q^2}\right),
\label{eq:K}
\end{equation}
with $ q_{\varepsilon} = \sqrt{\varepsilon k^2-k_z^2} $. Consequently, the GF elements~\eqref{eq:GT11},~\eqref{eq:GT22},~\eqref{eq:GT33}, and~\eqref{eq:GT13} change to
\begin{subequations}
\begin{alignat}{1}
\notag & G_{\mathbb{T}11} = \frac{i}{4\pi}\int_0^{\infty}dk_z\frac{k_z^2}{k^2}H^2_1(qr)T_{0,k_z}^{NN}\cos(k_zd)\\
& \ \ \ \ \ \ \ \ \ \ +\frac{i}{2\pi}\sum_{n=1}^{\infty}\int_0^{\infty}dk_z\left[\frac{n^2}{(qr)^2}H^2_n(qr)T_{n,k_z}^{MM} + 2\frac{nk_z}{kqr}H_n(qr)H'_n(qr)T_{n,k_z}^{MN} + \frac{k_z^2}{k^2}\left[H'_n(qr)\right]^2T_{n,k_z}^{NN}\right]\cos(k_zd),\label{eq:GT11e}\\
\notag & G_{\mathbb{T}22} = \frac{i}{4\pi}\int_0^{\infty}dk_zH^2_1(qr)T_{0,k_z}^{MM}\cos(k_zd)\\
& \ \ \ \ \ \ \ \ \ \ +\frac{i}{2\pi}\sum_{n=1}^{\infty}\int_0^{\infty}dk_z\left[[H'_n(qr)]^2T_{n,k_z}^{MM} + 2\frac{nk_z}{kqr}H_n(qr)H'_n(qr)T_{n,k_z}^{MN} +\frac{n^2k_z^2}{k^2(qr)^2}H^2_n(qr)T_{n,k_z}^{NN}\right]\cos(k_zd),\label{eq:GT22e}\\
& G_{\mathbb{T}33} = \frac{i}{4\pi}\int_0^{\infty}dk_z\frac{q^2}{k^2}H^2_0(qr)T_{0,k_z}^{NN}\cos(k_zd) + \frac{i}{2\pi}\sum_{n=1}^{\infty}\int_0^{\infty}dk_z\frac{q^2}{k^2}H^2_n(qr)T_{n,k_z}^{NN}\cos(k_zd),\label{eq:GT33e}\\
\notag & G_{\mathbb{T}13} = -\frac{i}{4\pi}\int_0^{\infty}dk_z\frac{qk_z}{k^2}H_0(qr)H_1(qr)T_{0,k_z}^{NN}\sin(k_zd)\\
& \ \ \ \ \ \ \ \ \ \ +\frac{i}{2\pi}\sum_{n=1}^{\infty}\int_0^{\infty}dk_z\left[\frac{n}{kr}H^2_n(qr)T_{n,k_z}^{MN} + \frac{qk_z}{k^2}H_n(qr)H'_n(qr)T_{n,k_z}^{NN}\right]\sin(k_zd),\label{eq:GT13e}
\end{alignat}
\end{subequations} 
with $ T_{n,k_z}^{MM} $, $ T_{n,k_z}^{NN} $, and $ T_{n,k_z}^{MN} $ given by Eqs.~\eqref{eq:TMMe},~\eqref{eq:TNNe}, and~\eqref{eq:TMNe}, respectively. In order to model gold cylinder, we use the following dielectric function (the Drude model for gold)~\cite{Ordal1983}:
\begin{equation}
\varepsilon_{\rm Au}(\omega) = 1-\frac{\omega_p^2}{\omega(\omega+i\omega_{\tau})},
\label{eq:epsilon_gold}
\end{equation}
with $ \omega_p = 1.37\times 10^{16} \ {\rm rad} \ {\rm s}^{-1} $ and $ \omega_{\tau} = 4.06\times 10^{13} \ {\rm rad} \ {\rm s}^{-1} $.

\subsubsection{Heat transfer}
Figure~\ref{fig:HT_gold} in the main text shows the HT in the presence of a gold cylinder. It has a very similar behavior compared to the perfect metal case (with even larger values of the HT, especially for small $ R $) until the decay length $ l_{\textrm{Au}} $, where the logarithmic dependence turns to a fast decay, finally converging to the vacuum HT. This decay length grows with $ R $. More precisely, based on the exponential form of the fast decay, we define $ l_{\textrm{Au}} $ as the distance where the gold HT is $ e $ times smaller than the corresponding perfect metal HT. The inset of Fig.~\ref{fig:HT_gold} in the main text shows $ l_{\textrm{Au}} $ as a function of $ R $. Although the Minsk-G\"ottingen experiment discussed in the main text would fail for a gold cylinder, still a very large $ l_{\textrm{Au}} $ (compared to other system length scales) is possible (e.g., $ l_{\textrm{Au}} \approx 4 \ \textrm{mm} $ for $ R = 1 \ \mu\textrm{m} $). In that respect, it would be interesting to investigate the HT in the presence of a superconducting cylinder, where an even larger decay length may be expected.


\end{document}